\documentclass[aps,pre,showpacs,onecolumn,superscriptaddress,amssymb]{revtex4}
\usepackage[dvips]{graphicx}

\newcommand{\Tr}{\mathrm{Tr}}
\newcommand{\bTr}{\mathrm{bTr}}
\newcommand{\D}{\mathrm{D}}

\newcommand{\cl}{\mathrm{cl.}}
\newcommand{\rmat}{\mathrm{r. m.}}

\newcommand{\im}{\mathrm{Im}}

\newcommand{\diag}{\mathrm{diag}}

\begin{document}

\title{Addition of Free Unitary Random Matrices}

\author{Andrzej T. G\"orlich}
\email{atg@th.if.uj.edu.pl}
\affiliation{M. Smoluchowski Institute of Physics and Mark Kac Center for Complex Systems Research, Jagellonian University, PL--30--059 Cracow, Poland}
\author{Andrzej Jarosz}
\email{jarosz@nbi.dk}
\affiliation{Niels Bohr Institute, DK--2100 Copenhagen, Denmark}
\affiliation{M. Smoluchowski Institute of Physics and Mark Kac Center for Complex Systems Research, Jagellonian University, PL--30--059 Cracow, Poland}

\date{\today}

\setlength{\parindent}{2ex}
\setlength{\parskip}{1ex plus 0.5ex minus 0.2ex}


\begin{abstract}
We consider a new class of non--Hermitian random matrices, namely the ones which have the form of sums of freely independent terms involving unitary matrices. To deal with them, we exploit the recently developed quaternion technique. After having derived some general identities describing additive properties of unitary matrices, we solve three particular models: CUE plus CUE, CUE plus \ldots plus CUE (\emph{i. e.} the sum of an arbitrary number of CUE matrices), and CUE plus GUE. By solution of a given model we mean here to calculate the borderline of the eigenvalues' two--dimensional domain, as well as the eigenvalues' density function inside the domain. We conf\mbox{}irm numerically all the results, obtaining very good agreement.

\end{abstract}



\maketitle


\section{Introduction}
\label{s:Introduction}


\subsection{Preface}

Recently there is growing interest in non--Hermitian random matrix models. One f\mbox{}inds them twofold interesting. F\mbox{}irst, non--Hermitian random matrix models are fascinating from the mathematical point of view, because their spectra cover two--dimensional and often multiple--connected supports on the whole complex plane, on the contrary to Hermitian ensembles, whose real eigenvalues form cuts on the real axis. This is enough to make most of the methods of Hermitian random matrix theory fail. Second, they are ubiquitous in dif\mbox{}ferent f\mbox{}ields of physics and interdisciplinary sciences. It is enough to mention open chaotic scattering~\cite{HAAKE}, spectral properties of Euclidian Dirac operators in the presence of chemical potential~\cite{CHEM}, the CP--violating angle $\theta$ in QCD~\cite{THETA}, non--Hermitian generalizations of the Anderson localization in mesoscopic systems~\cite{HATANO}, modeling of chemical transitions in dissipative systems~\cite{EWA}, matrix generalizations of multiplicative dif\mbox{}fusion processes~\cite{OURRECENT} or evolution of spectral curves of non--Hermitian ensembles in the context of the growth problem~\cite{TEODOR}.

Recently, a new technique of dealing with non--Hermitian random matrix models in the large--$N$ limit has been established~\cite{JAROSZNOWAK}, under the name of \emph{quaternion approach}, based on free random variables theory~\cite{FRV}. It allows to solve non--Hermitian models that acquire the form of sums of ``independent'' (free, see below) random matrices, in a simple algebraic way.

In this paper we shall concentrate on a new class of non--Hermitian sums of random matrices and apply the quaternion technique to solve them. These will be models having \emph{unitary} matrices as summands. Unitary random matrices f\mbox{}ind applications \emph{e. g.} in chaotic scattering~\cite{BLUMEL}, conductance in mesoscopic systems~\cite{BEENAKKER} or periodically driven quantum systems~\cite{HAAKE2}. The most important ensembles of this type are the \emph{Dyson models}~\cite{DYSON}, \emph{i. e.} the so--called COE, CUE and CSE; see also~\cite{GENERALHER} for a detailed discussion. We shall pick CUE as our basic example. For a variety of issues concerning unitary models we refer to~\cite{ZYCZKOWSKI}.

In other words, we shall try to \emph{add unitary random matrices}. This may seem quite odd because what one usually does with unitary matrices is to multiply them since the product of unitary matrices remains unitary, which is not the case when adding them. Moreover, any possible applications of such sums of unitary random matrices are still unclear. Nevertheless, we think that the problem is interesting at least from the mathematical point of view and its solution reveals how easy it is to treat non--Hermitian sums of random matrices using the quaternion technique, even if these sums appear very involved.

The rest of sec.~\ref{s:Introduction} is devoted to short summary of the quaternion technique. In sec.~\ref{s:QuaternionFormalismforUnitaryRandomMatrices} we apply the quaternion method to the case of sums of free unitary random matrices, deriving a number of useful formulae. Sec.~\ref{s:CUEPlusCUEModel},~\ref{s:CUEPlusldotsPlusCUEModel} and~\ref{s:CUEPlusGUEModel} contain solutions of three examples: the sum of two free CUE matrices, the sum of an arbitrary (and eventually also inf\mbox{}inite) number of free CUE matrices and the sum of the free CUE and GUE matrices. All the results are conf\mbox{}irmed by numerical computations. Finally, sec.~\ref{s:SummaryAndProspects} summarizes the paper and points to some future possibilities of exploring the subject.


\subsection{Summary of Quaternion Approach}

Let us here very brief\mbox{}ly recall the foundations of the quaternion approach; details and further references can be found in~\cite{JAROSZNOWAK}.

\begin{description}

\item[Free random variables.] The theoretical background of the quaternion approach contains the \emph{free random variables} calculus (FRV)~\cite{FRV}, which is a generalization of classical probability calculus to the case where random variables become non--commutative. In such a non--commutative setting it is possible to def\mbox{}ine a pertinent extension of the notion of independence, called \emph{freeness}, which allows to lift the ideas known from classical probability and built upon the concept of independence onto the level of non--commutative random variables; these ideas are \emph{e. g.} additive and multiplicative convolution, central limit theorems, \emph{etc.}

We shall deal with one of them, the problem of adding two free non--commutative random variables, $a_{1}$ and $a_{2}$. On the classical level this is readily solved by introducing the \emph{characteristic function} $g_{a} ( z ) \equiv \sum_{n \geq 0} \frac{\langle a^{n} \rangle}{n!} z^{n} = \langle e^{z a} \rangle$, which collects all the moments in a single generating function; and its logarithm satisf\mbox{}ies the \emph{addition law}, $\ln g_{a_{1} + a_{2}} ( z ) = \ln g_{a_{1}} ( z ) + \ln g_{a_{2}} ( z )$, for two independent commutative random variables $a_{1}$ and $a_{2}$. Hence in classical probability it is just logarithm that makes an additive function from the generating function of the moments. On the non--commutative level the task is much more involved but the f\mbox{}inal algorithm is analogous: The moments are gathered in another generating function, called the \emph{Green's function},
\begin{equation}
G_{a} ( z ) \equiv \sum_{n \geq 0} \frac{\langle a^{n} \rangle}{z^{n + 1}} = \left< \frac{1}{z - a} \right> ,
\end{equation}
and what needs to be done in order to obtain an additive function under the addition of two free non--commutative random variables is not to take logarithm but to invert it functionally,
\begin{equation}
G_{a} ( B_{a} ( z ) ) = B_{a} ( G_{a} ( z ) ) = z ,
\end{equation}
where this new object is called the \emph{Blue's function} and obeys the \emph{addition law},
\begin{equation}
B_{a_{1} + a_{2}} ( z ) = B_{a_{1}} ( z ) + B_{a_{2}} ( z ) - \frac{1}{z} .
\end{equation}

\item[Hermitian random matrix theory.] Random matrix theory (RMT) in the large--$N$ limit, which deals with large $N \times N$ random matrices, is a particular instance of the FRV calculus, with the following expectation value map,
\begin{equation}
\langle X \rangle_{\rmat} \equiv \left< \frac{1}{N} \Tr X \right>_{\cl} ,
\end{equation}
where $\langle \ldots \rangle_{\cl}$ is the classical expectation value. We shall focus on the eigenvalues of a given random matrix model. If we restrict to Hermitian matrices, call $H$ one of them, the eigenvalues $\lambda_{1} , \ldots , \lambda_{N}$ are \emph{real} and their \emph{density} is def\mbox{}ined through
\begin{equation}
\rho_{H} ( \lambda ) = \frac{1}{N} \left< \sum_{i = 1}^{N} \delta ( \lambda - \lambda_{i} ) \right>_{\cl} ;
\end{equation}
this is the crucial object that we aim to f\mbox{}ind. It turns out that the same Green's function which is so useful in the addition problem in general FRV, and which here acquires the form
\begin{equation} \label{eq:ComplexGreen}
G_{H} ( z ) = \frac{1}{N} \left< \Tr \frac{1}{z 1_{N} - H} \right>_{\cl} ,
\end{equation}
($1_{N}$ is the $N \times N$ unit matrix) serves perfectly also to investigate the eigenvalues' density: it is a meromorphic function with poles at the eigenvalues on the real axis, which in the large--$N$ limit become cuts, and its behaviour close to these cuts provides the desired eigenvalues' density,
\begin{equation}
\rho_{H} ( \lambda ) = - \frac{1}{\pi} \lim_{\epsilon \to 0^{+}} \im G_{H} ( \lambda + i \epsilon ) .
\end{equation}

\item[Non--Hermitian random matrix theory.] When we release the symmetry requirement that the matrices are Hermitian, the theory complicates signif\mbox{}icantly, which is mainly due to the fact that eigenvalues depart from the real line and become \emph{complex} in general, forming in the large--$N$ limit \emph{two--dimensional domains} instead of cuts. This deprives the Green's function of its usefulness as it is singular inside the eigenvalues' domains and can no longer provide the eigenvalues' density $\rho_{X} ( z , \bar{z} )$ of a non--Hermitian random matrix $X$. The idea is to \emph{regularize} the Green's function,
\begin{equation}
G_{X} ( z , \bar{z} ) \equiv \frac{1}{N} \left< \Tr \frac{\bar{z} 1_{N} - X^{\dagger}} {( z 1_{N} - X ) ( \bar{z} 1_{N} - X^{\dagger} ) + \epsilon^{2} 1_{N}} \right>_{\cl} ,
\end{equation}
where we put an additional factor of $( \bar{z} 1_{N} - X^{\dagger} )$ both in the numerator and denominator, and moreover the denominator is regularized by $\epsilon^{2} 1_{N}$ with $\epsilon \to 0^{+}$ (we shall skip the limit symbol henceforth). This is called the \emph{non--holomorphic Green's function}. It is def\mbox{}ined everywhere, also inside the eigenvalues' domains, as the regularized denominator is positive def\mbox{}inite on the whole complex plane; however, it is non--holomorphic. Not only have we now gained access to eigenvalues' domains but we can also reproduce the eigenvalues' density from the non--holomorphic Green's function via the simple formula
\begin{equation} \label{eq:EigenvaluesDensity}
\rho_{X} ( z , \bar{z} ) = \frac{1}{\pi} \partial_{\bar{z}} G_{X} ( z , \bar{z} ) .
\end{equation}

Even though the non--holomorphic Green's function is so meaningful, it seems to be very involved to evaluate due to the quadratic structure in its denominator instead of the linear one for the complex Green's function. This obstacle can be overcome by introducing yet a new object, the \emph{matrix--valued Green's function}, which is a $2 \times 2$ matrix def\mbox{}ined as
\begin{equation} \label{eq:MatrixValuedGreen}
\mathcal{G}_{X} ( z , \bar{z} ) \equiv \frac{1}{N} \left< \bTr \left( \begin{array}{cc} z 1_{N} - X & i \epsilon 1_{N} \\ i \epsilon 1_{N} & \bar{z} 1_{N} - X^{\dagger} \end{array} \right)^{- 1}_{2 N \times 2 N} \right>_{\cl} = \frac{1}{N} \left< \bTr \frac{1}{Z_{\epsilon} \otimes 1_{N} - X^{\D}} \right>_{\cl} ,
\end{equation}
where the \emph{block--trace} operation is
\begin{equation}
\bTr \left( \begin{array}{cc} A & B \\ C & D \end{array} \right)_{2 N \times 2 N} \equiv \left( \begin{array}{cc} \Tr A & \Tr B \\ \Tr C & \Tr D \end{array} \right)_{2 \times 2} ,
\end{equation}
and for short
\begin{equation}
Z_{\epsilon} \equiv \left( \begin{array}{cc} z & i \epsilon \\ i \epsilon & \bar{z} \end{array} \right)_{2 \times 2} ,
\end{equation}
\begin{equation}
X^{\D} \equiv \left( \begin{array}{cc} X & \\ & X^{\dagger} \end{array} \right)_{2 N \times 2 N} .
\end{equation}
This matrix looks as the usual complex Green's function but lifted to the level of $2 \times 2$ matrices and this similarity implies that some methods of Hermitian RMT used to derive the complex Green's function can be applied also in non--Hermitian RMT for the matrix--valued Green's function. Moreover, its $11$--element equals precisely the non--holomorphic Green's function,
\begin{equation}
G_{X} ( z , \bar{z} ) = [ \mathcal{G}_{X} ( z , \bar{z} ) ]_{11} ,
\end{equation}
which in this way becomes accessible.

The rest of the matrix--valued Green's function, \emph{i. e.} its of\mbox{}f--diagonal elements, may seem to carry no information, but actually they are important as well: their product
\begin{equation}
C_{X} ( z , \bar{z} ) \equiv [ \mathcal{G}_{X} ( z , \bar{z} ) ]_{1 2} [ \mathcal{G}_{X} ( z , \bar{z} ) ]_{2 1} ,
\end{equation}
evaluates the \emph{correlator between left and right eigenvectors of $X$},
\begin{equation}
\frac{1}{N} \left< \sum_{i = 1}^{N} ( L_{i} | L_{i} ) ( R_{i} | R_{i} ) \delta^{(2)} ( z - \lambda_{i} ) \right>_{\cl} = - \frac{1}{\pi} C_{X} ( z , \bar{z} ) .
\end{equation}
We shall exploit one of the properties of this object, namely the fact that it vanishes on the \emph{borderline} of the eigenvalues' domains and remains zero outside: hence a very important geometric characteristics of a given non--Hermitian model, the equation of the borderline, is readily obtained from the correlator,
\begin{equation} \label{eq:Borderline}
C_{X} ( z , \bar{z} ) = 0 .
\end{equation}
Therefore we shall be interested in the whole matrix--valued Green's function as it provides us with both the eigenvalues' density and the shape of the eigenvalues' domains.

\item[Quaternion formalism.] This new approach makes use of insights from both the FRV calculus and non--Hermitian RMT. The matrix--valued Green's function is generalized to a quaternion function of a quaternion variable, where we recall that a quaternion is a $2 \times 2$ matrix def\mbox{}ined by two complex numbers $a$ and $b$ via
\begin{equation}
Q = \left( \begin{array}{cc} a & i \bar{b} \\ i b & \bar{a} \end{array} \right)_{2 \times 2} ,
\end{equation}
by replacing $Z_{\epsilon}$ with an arbitrary quaternion $Q$, which gives the \emph{quaternion Green's function},
\begin{equation} \label{eq:QuaternionGreen}
\mathcal{G}_{X} ( Q ) \equiv \frac{1}{N} \left< \bTr \frac{1}{Q \otimes 1_{N} - X^{\mathrm{D}}} \right>_{\mathrm{cl}} .
\end{equation}
For $Q = Z_{\epsilon}$ we regain the former meaning together with its usefulness, but the complete functional dependence on $Q$ proves to be crucial. Indeed, it allows, following the FRV analogy, to invert $\mathcal{G}_{X} ( Q )$ functionally to get the \emph{quaternion Blue's function},
\begin{equation} \label{eq:QuaternionBlue}
\mathcal{G}_{X} ( \mathcal{B}_{X} ( Q ) ) = \mathcal{B}_{X} ( \mathcal{G}_{X} ( Q ) ) = Q .
\end{equation}
This new object might have turned out to be spurious as it generalizes just some notion from non--Hermitian RMT, whereas the Blue's function is a FRV concept. But it may be shown that in a general setting of the so--called \emph{FRV calculus with amalgamation}, which is an extension of usual FRV, the quaternion Blue's function plays a fully analogous role to the usual Blue's function, namely it obeys the \emph{quaternion addition law},
\begin{equation} \label{eq:QuaternionAdditionLaw}
\mathcal{B}_{X_{1} + X_{2}} ( Q ) = \mathcal{B}_{X_{1}} ( Q ) + \mathcal{B}_{X_{2}} ( Q ) - \frac{1}{Q} ,
\end{equation}
for two free random matrices $X_{1}$ and $X_{2}$, Hermitian or non--Hermitian. Therefore the addition problem in non--Hermitian RMT is solved: it relies on the knowledge of the quaternion Green's functions of both summands, $X_{1}$ and $X_{2}$, inverting them functionally, exploiting the quaternion addition law to get the quaternion Blue's function of $X_{1} + X_{2}$, and f\mbox{}inally inverting it functionally at the point of $Q = \diag ( z , \bar{z} )$ to obtain the matrix--valued Green's function of $X_{1} + X_{2}$, \emph{i. e.} solving
\begin{equation} \label{eq:Inversion2}
\mathcal{B}_{X_{1} + X_{2}} \left( \mathcal{G}_{X_{1} + X_{2}} ( z , \bar{z} ) \right) = \diag ( z , \bar{z} ) ,
\end{equation}
where we can set $\epsilon = 0$ as the regularization procedure happens to be entirely encoded in the functional inversion. This technique is called the \emph{quaternion addition algorithm}.

\end{description}


\section{Quaternion Formalism for Unitary Random Matrices}
\label{s:QuaternionFormalismforUnitaryRandomMatrices}


\subsection{Introduction}

Let us start with the following observation, which continues the considerations from the basic paper~\cite{JAROSZNOWAK}. We have seen that it is the quaternion Green's function $\mathcal{G}_{X} ( Q )$ (\ref{eq:QuaternionGreen}) which is the basic object to deal with a non--Hermitian random matrix model $X$ within the quaternion formalism. For instance, to apply the quaternion addition algorithm we need to know the quaternion Green's functions of both summands. Hence it becomes crucial to have a means of computing this object for some classes of random matrices. Now it is observed that there are cases where the quaternion Green's function can be quite easily calculated explicitly. More precisely, there are cases where it can be reduced to the complex Green's function $G_{X} ( z )$ (\ref{eq:ComplexGreen}),
\begin{equation} \label{eq:Passage}
G_{X} ( z ) \qquad \longrightarrow \qquad \mathcal{G}_{X} ( Q ) ;
\end{equation}
and this passage is called the \emph{Hermitization procedure}. In other words, there can be identif\mbox{}ied instances of $X$ such that the whole information contained in the quaternion Green's function follows from a much simpler object, the complex Green's function.

To see how such an idea could arise, let us note that $\mathcal{G}_{X} ( Q )$ is the averaged block trace of the inversion of a $2 N \times 2 N$ matrix, and let us simply perform this matrix inversion,
\begin{displaymath}
\mathcal{G}_{X} ( Q ) = \frac{1}{N} \left< \bTr \left( Q^{\mathrm{U}} - X^{\mathrm{D}} \right)^{- 1} \right> = \frac{1}{N} \left< \bTr \left(\begin{array}{cc} c 1_{N} - X & i \bar{d} 1_{N} \\ i d 1_{N} & \bar{c} 1_{N} - X^{\dagger} \end{array} \right)_{2 N \times 2 N}^{- 1} \right> =
\end{displaymath}
\begin{equation} \label{eq:QuaternionGreenExplicit}
= \frac{1}{N} \left< \bTr \left( \begin{array}{cc} \frac{\bar{c} 1_{N} - X^{\dagger}}{X X^{\dagger} - c X^{\dagger} - \bar{c} X + ( | c |^{2} + | d |^{2} ) 1_{N}} & \frac{- i \bar{d}}{X X^{\dagger} - c X^{\dagger} - \bar{c} X + ( | c |^{2} + | d |^{2} ) 1_{N}} \\ \frac{- i d}{X^{\dagger} X - c X^{\dagger} - \bar{c} X + ( | c |^{2} + | d |^{2} ) 1_{N}} & \frac{c 1_{N} - X}{X^{\dagger} X - c X^{\dagger} - \bar{c} X + ( | c |^{2} + | d |^{2} ) 1_{N}} \end{array} \right)_{2 N \times 2 N} \right> ,
\end{equation}
where now we denote the quaternion $Q$ by
\begin{equation} \label{eq:Quaternion}
Q \equiv \left( \begin{array}{cc} c & i \bar{d} \\ i d & \bar{c} \end{array} \right)_{2 \times 2} ,
\end{equation}
to agree with our later conventions.

Now every block of the above $2 N \times 2 N$ matrix is seen to be a rational function of $X$ and $X^{\dagger}$. The denominators here are quadratic, which is in contrast with the linear denominator featuring in the complex Green's function (\ref{eq:ComplexGreen}), and this mainly hinders ef\mbox{}f\mbox{}icient computation of $\mathcal{G}_{X} ( Q )$. However, if there is a symmetry restriction on the matrix $X$ which assumes the following form,
\begin{equation} \label{eq:Rational}
X^{\dagger} = \textrm{rational function of $X$,}
\end{equation}
then every block of the matrix in question becomes actually a rational function only of $X$. If so, it can be expanded in simple fractions, \emph{i. e.} functions of the form $1 / ( s 1_{N} - X )$, for some complex numbers $s$; this is nothing but the structure appearing in $G_{X} ( s )$, hence the full quaternion Green's function seems likely to be expressed solely in terms of the complex Green's function, thus proving the Hermitization procedure (\ref{eq:Passage}) for $X$.

The symmetry constraint (\ref{eq:Rational}), even though at f\mbox{}irst it looks quite fanciful, is not of minor importance. There are at least its two important realizations,
\begin{equation}
H^{\dagger} = H \qquad \textrm{(Hermitian)}, \qquad U^{\dagger} = U^{- 1} \qquad \textrm{(unitary)} .
\end{equation}
In both these cases, an appropriate Hermitization procedure should be applicable, thus reducing a complicated quaternion object to a much simpler comlpex one.

In the Hermitian case, $X = H$, the quaternion Green's function has been Hermitized in the basic paper~\cite{JAROSZNOWAK}, and let us just quote the resulting formula for further comparison,
\begin{equation} \label{eq:QuaternionGreenForH}
\mathcal{G}_{H} ( Q ) = \gamma_{H} ( q , \bar{q} ) 1_{2} - \gamma^{\prime}_{H} ( q , \bar{q} ) Q^{\dagger} ,
\end{equation}
where $\gamma_{H}$ and $\gamma^{\prime}_{H}$ are two scalar functions depending only on the eigenvalues $q , \bar{q}$ of $Q$, and given by
\begin{equation} \label{eq:GammaH}
\gamma_{H} ( q , \bar{q} ) \equiv \frac{q G_{H} ( q ) - \bar{q} G_{H} ( \bar{q} )}{q - \bar{q}} , \qquad \gamma^{\prime}_{H} ( q , \bar{q} ) \equiv \frac{G_{H} ( q ) - G_{H} ( \bar{q} )}{q - \bar{q}} .
\end{equation}
This allows to have the full quaternion object $\mathcal{G}_{H} ( Q )$ once the complex one, $G_{H} ( z )$, is known.


\subsection{Quaternion Green's Function for Unitary Random Matrix}

The same idea can be now applied also in the unitary case, $X = U$. The constraint $U^{\dagger} = U^{- 1}$ shall be substituted to (\ref{eq:QuaternionGreenExplicit}), and the fractions expanded in simple fractions. Due to the quaternion structure of the result, it is suf\mbox{}f\mbox{}icient to write only \emph{e. g.} the left--upper and left--lower elements, the other (depicted by stars) simply follow from these two via appropriate complex conjugation, compare (\ref{eq:Quaternion}),
\begin{displaymath}
\mathcal{G}_{U} ( Q ) = \frac{1}{N} \left< \bTr \left( \begin{array}{cc} \frac{\bar{c} 1_{N} - U^{- 1}}{g 1_{N} - c U^{- 1} - \bar{c} U} & \frac{- i \bar{d}}{g 1_{N} - c U^{- 1} - \bar{c} U} \\ \frac{- i d}{g 1_{N} - c U^{- 1} - \bar{c} U} & \frac{c 1_{N} - U}{g 1_{N} - c U^{- 1} - \bar{c} U} \end{array} \right)_{2 N \times 2 N} \right> = \frac{1}{N} \left< \bTr \frac{1}{u_{1} - u_{2}} \left( \begin{array}{cc} \frac{u_{1} - \frac{1}{\bar{c}}}{u_{1} 1_{N} - U} + \frac{- u_{2} + \frac{1}{\bar{c}}}{u_{2} 1_{N} - U} & \bigstar \\ \frac{- \frac{i d}{\bar{c}} u_{1}}{u_{1} 1_{N} - U} + \frac{\frac{i d}{\bar{c}} u_{2}}{u_{2} 1_{N} - U} & \bigstar \end{array} \right)_{2 N \times 2 N} \right> =
\end{displaymath}
\begin{equation} \label{eq:QuaternionGreenForU}
= \left( \begin{array}{cc} \gamma_{U} ( Q ) - \frac{1}{\bar{c}} \gamma^{\prime}_{U} ( Q ) & \bigstar \\ \frac{- i d}{\bar{c}} \gamma_{U} ( Q ) & \bigstar \end{array} \right)_{2 \times 2} ,
\end{equation}
where we introduce the following notation,
\begin{equation} \label{eq:GammaU}
\gamma_{U} ( Q ) \equiv \frac{u_{1} G_{U} ( u_{1} ) - u_{2} G_{U} ( u_{2} )}{u_{1} - u_{2}} , \qquad \gamma^{\prime}_{U} ( Q ) \equiv \frac{G_{U} ( u_{1} ) - G_{U} ( u_{2} )}{u_{1} - u_{2}} ,
\end{equation}
where $u_{1 , 2}$ are the solutions of the quadratic equation $\bar{c} u^{2} - g u + c = 0$,
\begin{equation}
u_{1 , 2} = \frac{1}{2 \bar{c}} \left( g \pm \sqrt{g^{2} - 4 | c |^{2}} \right) ,
\end{equation}
where for short
\begin{equation}
g \equiv | c |^{2} + | d |^{2} + 1 .
\end{equation}
In this simple way we are given the quaternion Green's function for any unitary random matrix model $U$, expressed via its complex Green's function.

Note similarities and dif\mbox{}ferences between (\ref{eq:QuaternionGreenForU}), (\ref{eq:GammaU}) in the unitary case and (\ref{eq:QuaternionGreenForH}), (\ref{eq:GammaH}) in the Hermitian one. In particular, in the Hermitian case the gamma functions depend on the eigenvalues $q , \bar{q}$ of $Q$, so they are invariant under similarity transformations of $Q$, whereas in the unitary case we have a more involved and non--rotationally invariant dependence of the gamma coef\mbox{}f\mbox{}icients on $u_{1 , 2}$.

Note that $\overline{u_{1}} u_{2} = u_{1} \overline{u_{2}} = 1$, hence we can denote
\begin{equation}
u \equiv u_{1} , \qquad \textrm{and consequently} \qquad u_{2} = \frac{1}{\bar{u}} ,
\end{equation}
which gives, due to the formula $\overline{G_{U} ( z )} = \frac{1}{\bar{z}} \left( 1 - \frac{1}{\bar{z}} G_{U} ( \frac{1}{\bar{z}} ) \right)$,
\begin{equation}
\gamma_{U} ( Q ) = \frac{u G_{U} ( u ) + \overline{u G_{U} ( u )} - 1}{u - \frac{1}{\bar{u}}} , \qquad \gamma^{\prime}_{U} ( Q ) = \frac{G_{U} ( u ) + \bar{u} \left( \overline{u G_{U} ( u )} - 1 \right)}{u - \frac{1}{\bar{u}}} .
\end{equation}
Thus we have expressed the gamma functions through the single complex variable $u$ instead of two mutually dependent variables $u_{1 , 2}$.

Let us note that we have obtained the basic expression (\ref{eq:QuaternionGreenForU}) under some slight assumptions,
\begin{itemize}
\item $c \neq 0$,
\item $u_{1} \neq u_{2}$, \emph{i. e.} $| c | \neq 1$ or $d \neq 0$.
\end{itemize}
{}From now on we have to consider these two special cases separately from the generic case. For $c = 0$ we get
\begin{equation} \label{eq:QuaternionGreenForUFora0}
\mathcal{G}_{U} \left( \left( \begin{array}{cc} 0 & i \bar{d} \\ i d & 0 \end{array} \right)_{2 \times 2} \right) = - \frac{1}{| d |^{2} + 1} \left( \begin{array}{cc} \overline{m_{U , 1}} & i \bar{d} \\ i d & m_{U , 1} \end{array} \right)_{2 \times 2} ,
\end{equation}
where $m_{U , 1} = \frac{1}{N} \langle \Tr U \rangle$ is the f\mbox{}irst moment of $U$. For $| c | = 1$ and $d = 0$ immediately
\begin{equation} \label{eq:QuaternionGreenForUFora1b0}
\mathcal{G}_{U} ( \mathrm{diag} ( c , \bar{c} ) ) = \mathrm{diag} ( G_{U} ( c ) , \overline{G_{U} ( c )} ) ;
\end{equation}
we have to pay attention when using this formula since here $| c | = 1$ and unitary random matrices have their eigenvalues exactly on the unit circle, so that $G_{U} ( c )$ is given by a divergent integral, which however makes sense as a certain limit.


\subsection{Example: Quaternion Green's Function for CUE}

Let us now consider a particular instance of a unitary random matrix, namely the \emph{circular unitary ensemble} (CUE), which means
\begin{equation}
P ( U ) = \mathrm{const} ,
\end{equation}
or in other words, the eigenvalues' density reads
\begin{equation}
\rho_{U} ( z , \bar{z} ) = \frac{1}{2 \pi} \qquad \textrm{on the unit circle.}
\end{equation}
The complex Green's function is derived easily,
\begin{displaymath}
G_{U} ( z ) = \frac{1}{2 \pi} \int_{0}^{2 \pi} \frac{\mathrm{d} \theta}{z - e^{i \theta}} = \frac{i}{2 \pi} \int_{0}^{2 \pi} \frac{\mathrm{d} e^{- i \theta}}{z e^{- i \theta} - 1} = - \frac{i}{2 \pi} \oint_{C ( 0 , 1 )} \frac{\mathrm{d} t}{z t - 1} = \left\{ \begin{array}{ll} \mathrm{Res}_{t = \frac{1}{z}} \frac{1}{z t - 1} , & \textrm{for $| z | > 1$} \\ 0 , & \textrm{for $| z | < 1$} \end{array} \right. =
\end{displaymath}
\begin{equation}
= \left\{ \begin{array}{ll} \frac{1}{z} , & \textrm{for $| z | > 1$} \\ 0 , & \textrm{for $| z | < 1$} \end{array} \right. .
\end{equation}

In order to calculate the quaternion Green's function, we have to f\mbox{}ind the gamma functions. Since there is always $| u | > 1$ due to $| u | > 1 \Leftrightarrow ( | c | - 1 )^{2} + | d |^{2} > 0$, we have
\begin{displaymath}
\gamma_{U} ( Q ) = \frac{\bar{c}}{\sqrt{g^{2} - 4 | c |^{2}}} , \qquad \gamma^{\prime}_{U} ( Q ) = - \frac{\bar{c}}{2 c} + \frac{g \bar{c}}{2 c \sqrt{g^{2} - 4 | c |^{2}}} ,
\end{displaymath}
therefore
\begin{equation}
[ \mathcal{G}_{U} ( Q ) ]_{1 1} = \frac{1}{2 c} + \frac{| c |^{2} - | d |^{2} - 1}{2 c \sqrt{g^{2} - 4 | c |^{2}}} , \qquad [ \mathcal{G}_{U} ( Q ) ]_{2 1} = - \frac{d}{\sqrt{g^{2} - 4 | c |^{2}}} .
\end{equation}
These constitute the quaternion Green's function for the CUE random matrix.

For the special case $c = 0$, since the f\mbox{}irst moment of the CUE random matrix vanishes,
\begin{equation}
\mathcal{G}_{U} \left( \left( \begin{array}{cc} 0 & i \bar{d} \\ i d & 0 \end{array} \right)_{2 \times 2} \right) = - \frac{1}{| d |^{2} + 1} \left( \begin{array}{cc} 0 & i \bar{d} \\ i d & 0 \end{array} \right)_{2 \times 2} .
\end{equation}


\subsection{Quaternion Blue's Function for Unitary Random Matrix}

We have seen that it is the quaternion Blue's function $\mathcal{B}_{X} ( Q )$ (\ref{eq:QuaternionBlue}), \emph{i. e.} the functional inverse of the quaternion Green's function, which is the basic object entering the quaternion addition law (\ref{eq:QuaternionAdditionLaw}), allowing us to add free random matrices.

Here we aim to invert functionally $\mathcal{G}_{U} ( Q )$ (\ref{eq:QuaternionGreenForU}). The resulting quaternion Blue's function is a quaternion denoted by
\begin{equation}
\mathcal{B}_{U} ( Q ) \equiv \left( \begin{array}{cc} c & i \bar{d} \\ i d & \bar{c} \end{array} \right)_{2 \times 2} ,
\end{equation}
where we return to the notation from sec.~\ref{s:Introduction} for the quaternion $Q$,
\begin{equation}
Q \equiv \left( \begin{array}{cc} a & i \bar{b} \\ i b & \bar{a} \end{array} \right)_{2 \times 2} .
\end{equation}
It is provided by solving the two equations
\begin{equation} \label{eq:QuaternionBlueForU}
\gamma_{U} ( \mathcal{B}_{U} ( Q ) ) - \frac{1}{\bar{c}} \gamma^{\prime}_{U} ( \mathcal{B}_{U} ( Q ) ) = a , \qquad - \frac{d}{\bar{c}} \gamma_{U} ( \mathcal{B}_{U} ( Q ) ) = b ,
\end{equation}
with two complex unknowns, $c$ and $d$, assuming the knowledge of $a$ and $b$, \emph{i. e.} the complex coef\mbox{}f\mbox{}icients of the quaternion $Q$. These equations are valid only for the generic case of $c \neq 0$ and ($| c | \neq 1$ or $d \neq 0$). We shall not solve them as we shall see that we actaully do not need an explicit solution, barely having equations is enough.

If $c = 0$ then the equations are
\begin{displaymath}
\overline{m_{U , 1}} = a , \qquad \frac{- d}{| d |^{2} + 1} = b ,
\end{displaymath}
which has a solution only while
\begin{equation}
0 < | b | < \frac{1}{2} , \qquad  a = \overline{m_{U , 1}} ,
\end{equation}
which is then of the form
\begin{equation} \label{eq:QuaternionBlueForUForc0}
c = 0 , \qquad d = - \frac{1 + \sqrt{1 - 4 | b |^{2}}}{2 \bar{b}} .
\end{equation}


\subsection{Example: Quaternion Blue's Function for CUE}

In the CUE case we know $G_{U} ( z )$ explicitly, so we can also explicitly write the equations (\ref{eq:QuaternionBlueForU}),
\begin{equation}
\frac{1}{2 c} + \frac{| c |^{2} - | d |^{2} - 1}{2 c \sqrt{g^{2} - 4 | c |^{2}}} = a , \qquad - \frac{d}{\sqrt{g^{2} - 4 | c |^{2}}} = b .
\end{equation}
This time we are also not going to solve this equations. The reason is that they will soon be used together with the quaternion addition law, which imposes some additional constraints on our unknowns $c$ and $d$, and it will turn out that it is easier to solve all these equations together.

Let us however simplify them a little. Let us assume f\mbox{}irst the generic case of $c \neq 0$, $a \neq 0$. We see from the f\mbox{}irst of these equations that
\begin{equation} \label{eq:QuaternionBlueForCUE1}
\alpha \equiv a c \in \mathbb{R} ,
\end{equation}
and we rewrite the second one as
\begin{equation} \label{eq:QuaternionBlueForCUE2}
d = - b \beta , \qquad \beta \equiv \sqrt{g^{2} - 4 | c |^{2}} \in \mathbb{R}_{+} .
\end{equation}
Hence we are left with the following two equations,
\begin{equation} \label{eq:QuaternionBlueForCUE3}
\beta = \sqrt{g^{2} - 4 \frac{\alpha^{2}}{| a |^{2}}} ,
\end{equation}
\begin{equation} \label{eq:QuaternionBlueForCUE4}
2 \alpha - 1 = \frac{\frac{\alpha^{2}}{| a |^{2}} - | b |^{2} \beta^{2} - 1}{\beta} ,
\end{equation}
where
\begin{equation} \label{eq:QuaternionBlueForCUE5}
g = \frac{\alpha^{2}}{| a |^{2}} + | b |^{2}\beta^{2} + 1 ,
\end{equation}
for two real unknowns, $\alpha$ and $\beta$.

If $c \neq 0$ it however may turn out that $a = 0$, and so the derivation does not hold. Instead, we get after a short calculation,
\begin{displaymath}
| c | = 1 , \qquad d = 0 ,
\end{displaymath}
which is impossible to be reached from $a = 0$ due to (\ref{eq:QuaternionGreenForUFora1b0}).

If $c = 0$, one must have $a = 0$ and $0 < | b | < \frac{1}{2}$, and then $d$ is given by (\ref{eq:QuaternionBlueForUForc0}).


\section{CUE Plus CUE Model}
\label{s:CUEPlusCUEModel}

We have established foundations of the quaternion approach in the case of unitary random matrices. Now let us move to a number of random matrix models which have the form of a sum of matrcies such as CUE or GUE.


\subsection{Introduction}

As the f\mbox{}irst example we consider the model
\begin{equation}
X = U_{1} + U_{2} ,
\end{equation}
where $U_{1 , 2}$ are two free CUE random matrices.

The quaternion addition law (\ref{eq:QuaternionAdditionLaw}) reads
\begin{displaymath}
\mathcal{B}_{U_{1} + U_{2}} ( Q ) = \mathcal{B}_{U_{1}} ( Q ) + \mathcal{B}_{U_{2}} ( Q ) - \frac{1}{Q} = 2 \mathcal{B}_{U} ( Q ) - \frac{1}{Q} ,
\end{displaymath}
\emph{i. e.} denoting
\begin{equation}
\mathcal{B}_{U_{1} + U_{2}} ( Q ) \equiv \left( \begin{array}{cc} C & i \bar{D} \\ i D & \bar{C} \end{array} \right)_{2 \times 2} ,
\end{equation}
and exploiting our previous notation,
\begin{equation}
C = 2 c - \frac{\bar{a}}{| a |^{2} + | b |^{2}} , \qquad D = 2 d + \frac{b}{| a |^{2}+| b |^{2}} .
\end{equation}

To solve the model (\emph{i. e.} to f\mbox{}ind the eigenvalues' density and the borderline of the eigenvalues' domain) we need to consider the matrix--valued Green's function (\ref{eq:MatrixValuedGreen}), \emph{i. e.} to solve the equation (\ref{eq:Inversion2}),
\begin{displaymath}
\mathcal{B}_{U_{1} + U_{2}} ( Q ) = \diag ( z , \bar{z} ) ,
\end{displaymath}
with respect to $Q$, where $z$ is a given complex number. This reads
\begin{equation} \label{eq:BasicCUE1}
z = 2 c - \frac{\bar{a}}{| a |^{2} + | b |^{2}} ,
\end{equation}
\begin{equation} \label{eq:BasicCUE2}
0 = 2 d + \frac{b}{| a |^{2} + | b |^{2}} .
\end{equation}
Our aim is now to solve these equations with respect to $a$ and $b$, where $c$ and $d$ are given by (\ref{eq:QuaternionBlueForCUE1}), (\ref{eq:QuaternionBlueForCUE2}), (\ref{eq:QuaternionBlueForCUE3}), (\ref{eq:QuaternionBlueForCUE4}), (\ref{eq:QuaternionBlueForCUE5}). Let us note that here $a \neq 0$, because otherwise $c = 0$ which would contradict (\ref{eq:BasicCUE1}).


\subsection{Solution}

Let us write down all the equations we have to solve. F\mbox{}irst, (\ref{eq:BasicCUE1}) multiplied by $a$ shows that
\begin{equation}
\gamma \equiv z a \in \mathbb{R} ,
\end{equation}
which in particular means that
\begin{equation}
| a |^{2} = \frac{\gamma^{2}}{| z |^{2}} ,
\end{equation}
so that (\ref{eq:BasicCUE1}) assumes the form
\begin{equation} \label{eq:Sol1}
\alpha = \frac{\gamma}{2} + \frac{\gamma^{2}}{2 \left( \gamma^{2} + | z |^{2} | b |^{2} \right)} .
\end{equation}
Second, (\ref{eq:BasicCUE2}) together with (\ref{eq:QuaternionBlueForCUE3}) acquires the form
\begin{equation} \label{eq:Sol2}
\beta = \frac{| z |^{2}}{2 \left( \gamma^{2} + | z |^{2} | b |^{2} \right)} .
\end{equation}
Third, let us recall (\ref{eq:QuaternionBlueForCUE3}), (\ref{eq:QuaternionBlueForCUE4}), (\ref{eq:QuaternionBlueForCUE5}) in the new setup, \emph{i. e.}
\begin{equation} \label{eq:Sol3}
\beta = \sqrt{g^{2} - 4 \frac{\alpha^{2} | z |^{2}}{\gamma^{2}}} ,
\end{equation}
\begin{equation} \label{eq:Sol4}
2 \alpha - 1 = \frac{\frac{\alpha^{2} | z |^{2}}{\gamma^{2}} - | b |^{2}\beta^{2} - 1}{\beta} ,
\end{equation}
\begin{equation} \label{eq:Sol5}
g = \frac{\alpha^{2} | z |^{2}}{\gamma^{2}} + | b |^{2} \beta^{2} + 1 .
\end{equation}
To sum up, we aim to solve f\mbox{}ive real equations, (\ref{eq:Sol1}), (\ref{eq:Sol2}), (\ref{eq:Sol3}), (\ref{eq:Sol4}), (\ref{eq:Sol5}), for f\mbox{}ive real unknowns, $\gamma$, $| b |^{2}$, $\alpha$, $\beta$, $g$, among which we are specif\mbox{}ically interested in $\gamma$ and $| b |^{2}$.

To exectute this plan, we substitute (\ref{eq:Sol1}) and (\ref{eq:Sol2}) into the remaining equations. First, into (\ref{eq:Sol4}), which simplif\mbox{}ies surprisingly into
\begin{displaymath}
\gamma^{2} = \frac{| z |^{2}}{4 - | z |^{2}} - | z |^{2} | b |^{2} .
\end{displaymath}
Second, to (\ref{eq:Sol5}),
\begin{displaymath}
g = 2 + 2 \gamma \left( 1 - \frac{| z |^{2}}{4} \right) .
\end{displaymath}
Third, to (\ref{eq:Sol3}), which gives a half of the solution,
\begin{equation}
| b |^{2} = \frac{2 \left( 2 - | z |^{2} \right)}{\left( 4 - | z |^{2} \right)^{2}} ,
\end{equation}
so that also the second half,
\begin{equation}
\gamma = \frac{| z |^{2}}{4 - | z |^{2}} .
\end{equation}

This can be rewritten in terms of the basic quantities,
\begin{equation}
G_{U_{1} + U_{2}} ( z , \bar{z} ) = \frac{\bar{z}}{4 - | z |^{2}} ,
\end{equation}
\begin{equation}
- C_{U_{1} + U_{2}} ( z , \bar{z} ) = \frac{2 \left( 2 - | z |^{2} \right)}{\left( 4 - | z |^{2} \right)^{2}} .
\end{equation}

Therefore, the eigenvalue's density (\ref{eq:EigenvaluesDensity}) is
\begin{equation} \label{eq:EigenvaluesDensityForCUE}
\rho_{U_{1} + U_{2}} ( z , \bar{z} ) = \frac{4}{\pi \left( 4 - | z |^{2} \right)^{2}} ,
\end{equation}
which can be easily checked to be normalized to $1$, whereas the borderline's equation (\ref{eq:Borderline}) reads
\begin{equation} \label{eq:BorderlineForCUE}
| z | = \sqrt{2} ,
\end{equation}
which means that the eigenvalues f\mbox{}ill the centered circle of radius $\sqrt{2}$.


\subsection{Numerical Conf\mbox{}irmation}

Let us numerically conf\mbox{}irm the above results. We have drawn $100$ pairs of $200 \times 200$ unitary matrices from the uniform distribution (CUE), added them, and diagonalized the sum.

The left f\mbox{}igure shows positions of these $20 000$ eigenvalues on the complex plane as well as the theoretical borderline (\ref{eq:BorderlineForCUE}). The right f\mbox{}igure presents the radial section of the eigenvalues' density plot (for it is circularly symmetric); the solid curve stands for the theoretical result (\ref{eq:EigenvaluesDensityForCUE}), and there is a numerical histogram under the curve.

\includegraphics[angle=-90,width=0.5\textwidth]{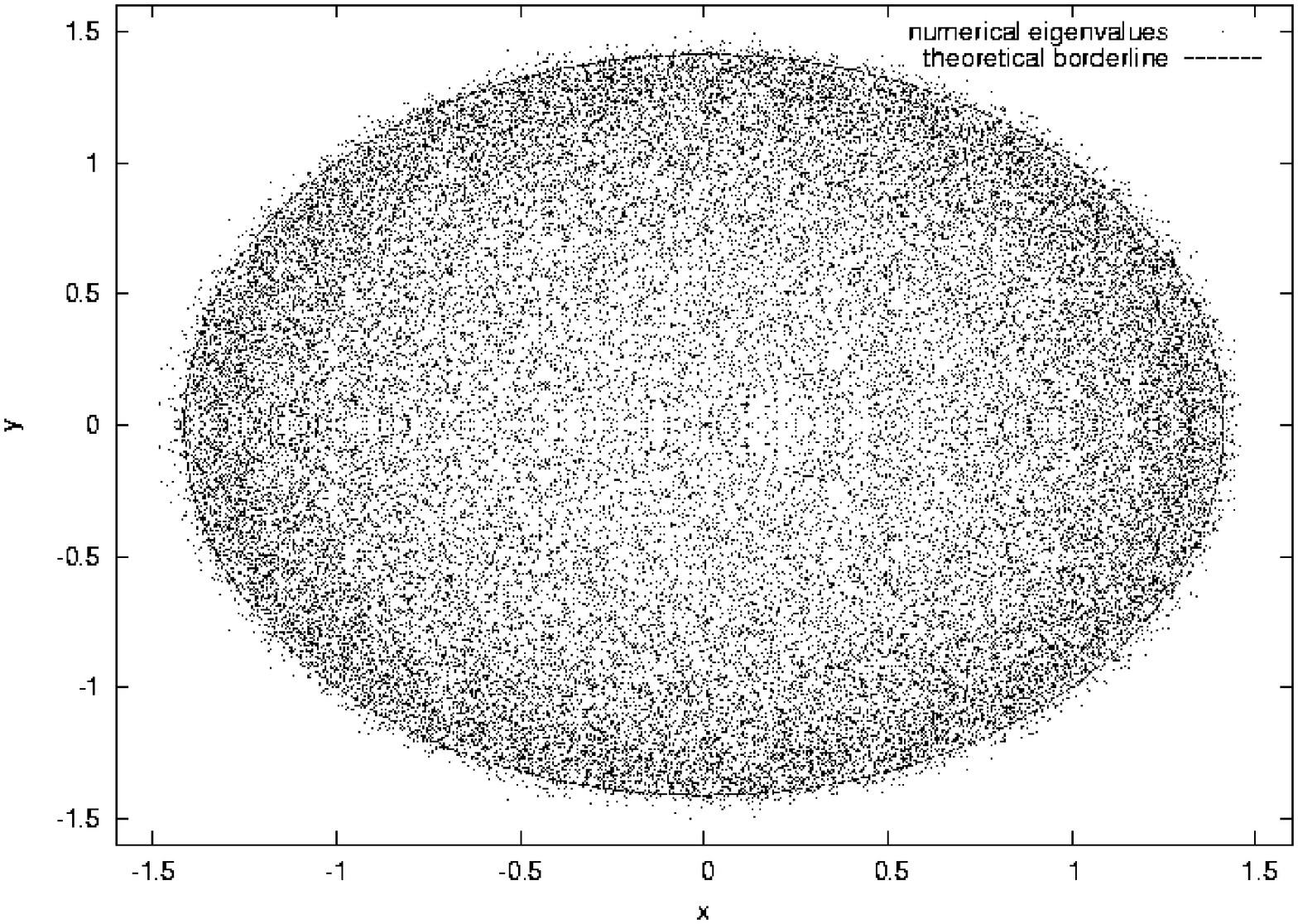}
\includegraphics[angle=-90,width=0.5\textwidth]{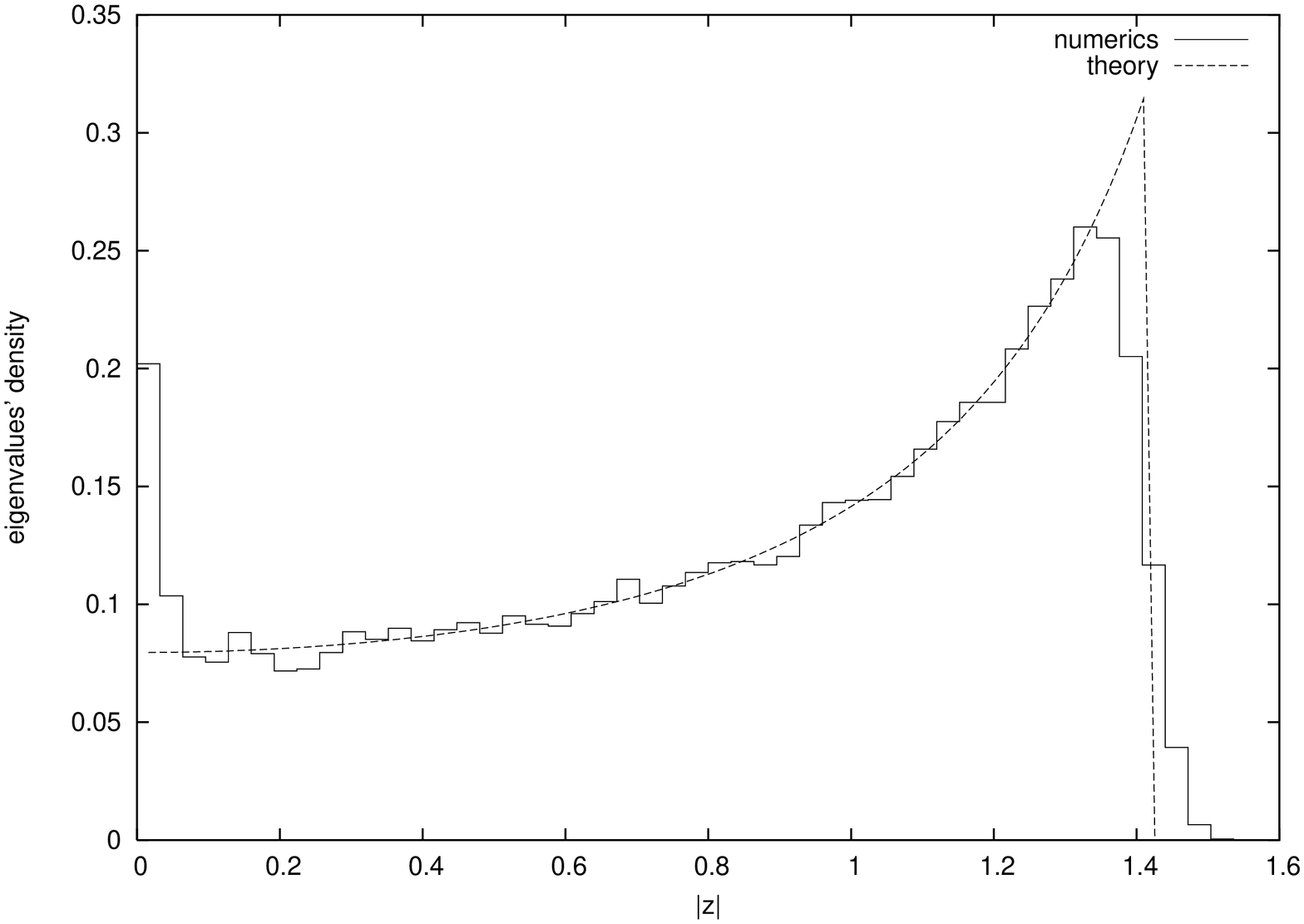}\\

We are convinced of the perfect agreement between theory and numerics.

Some slight deviations are due to a f\mbox{}inite ($N = 200$) size of matrices. In the origin there is also a strange peak visible, which we suspect to be a numerical artefact.


\section{CUE Plus \ldots Plus CUE Model}
\label{s:CUEPlusldotsPlusCUEModel}


\subsection{Introduction}

As the second model, let us consider a slight generalization of the CUE plus CUE model, namely the sum of $M \geq 2$ free CUE random matrices $U_{i}$,
\begin{equation}
X = U_{1} + \ldots + U_{M} .
\end{equation}
In particular, we shall eventually multiply $X$ by a constant depending on $M$ ($1 / \sqrt{M}$, specif\mbox{}ically) to ensure that the result has a correct behaviour for $M \to \infty$, and we shall investigate this limit. This will be free additive unitary dif\mbox{}fusion.

The quaternion addition law (\ref{eq:QuaternionAdditionLaw}) states that
\begin{displaymath}
\mathcal{B}_{U_{1} + \ldots + U_{M}} ( Q ) = M \mathcal{B}_{U} ( Q ) - \frac{M - 1}{Q} ,
\end{displaymath}
\emph{i. e.}
\begin{equation}
C = M c - ( M - 1 ) \frac{\bar{a}}{| a |^{2} + | b |^{2}} , \qquad D = M d + ( M - 1 ) \frac{b}{| a |^{2} + | b |^{2}} .
\end{equation}

Again, the basic equations are
\begin{equation} \label{eq:BasicMCUE1}
z = M c - ( M - 1 ) \frac{\bar{a}}{| a |^{2} + | b |^{2}} ,
\end{equation}
\begin{equation} \label{eq:BasicMCUE2}
0 = M d + ( M - 1 ) \frac{b}{| a |^{2} + | b |^{2}} .
\end{equation}
Again, only the generic case of $a \neq 0$ needs to be considered.


\subsection{Solution}

Let us write equations to solve. F\mbox{}irst, (\ref{eq:BasicMCUE1}) multiplied by $a$ gives
\begin{equation}
\gamma \equiv z a \in \mathbb{R} ,
\end{equation}
with (\ref{eq:BasicMCUE1}) being then
\begin{equation} \label{eq:2Sol1}
\alpha = \frac{1}{M} \gamma + \left( 1 - \frac{1}{M} \right) \frac{\gamma^{2}}{\gamma^{2} + | z |^{2} | b |^{2}} .
\end{equation}
Second, (\ref{eq:BasicMCUE2}) together with (\ref{eq:QuaternionBlueForCUE2}) assume the form
\begin{equation} \label{eq:2Sol2}
\beta = \left( 1 - \frac{1}{M} \right) \frac{| z |^{2}}{\gamma^{2} + | z |^{2} | b |^{2}} .
\end{equation}
Third, the basic equations (\ref{eq:QuaternionBlueForCUE3}), (\ref{eq:QuaternionBlueForCUE4}), (\ref{eq:QuaternionBlueForCUE5}) remain identical as in the CUE plus CUE case,
\begin{equation} \label{eq:2Sol3}
\beta = \sqrt{g^{2} - 4 \frac{\alpha^{2} | z |^{2}}{\gamma^{2}}} ,
\end{equation}
\begin{equation} \label{eq:2Sol4}
2 \alpha - 1 = \frac{\frac{\alpha^{2} | z |^{2}}{\gamma^{2}} - | b |^{2} \beta^{2} - 1}{\beta} ,
\end{equation}
\begin{equation} \label{eq:2Sol5}
g = \frac{\alpha^{2} | z |^{2}}{\gamma^{2}} + | b |^{2} \beta^{2} + 1 .
\end{equation}

We get thus to solve f\mbox{}ive real equations, (\ref{eq:2Sol1}), (\ref{eq:2Sol2}), (\ref{eq:2Sol3}), (\ref{eq:2Sol4}), (\ref{eq:2Sol5}) for f\mbox{}ive real variables, $\gamma$, $| b |^{2}$, $\alpha$, $\beta$, $g$, among which we are specif\mbox{}ically interested in $\gamma$ and $| b |^{2}$.

The equations are almost identical to those for the CUE plus CUE model. Hence proceeding analogously,
\begin{equation}
G_{U_{1} + \ldots + U_{M}} ( z , \bar{z} ) = \frac{\bar{z} ( M - 1 )}{M^{2} - | z |^{2}} ,
\end{equation}
\begin{equation}
- C_{U_{1} + \ldots + U_{M}} ( z , \bar{z} ) = \frac{M ( M - 1 ) \left( M - | z |^{2} \right)}{\left( M^{2} - | z |^{2} \right)^{2}} .
\end{equation}
For $M = 2$ we regain the results of the previous section.

The eigenvalues' density thus reads
\begin{equation} \label{eq:EigenvaluesDensityForMCUE}
\rho_{U_{1} + \ldots + U_{M}} ( z , \bar{z} ) = \frac{M^{2} ( M - 1 )}{\pi \left( M^{2} - | z |^{2} \right)^{2}} ,
\end{equation}
normalized to $1$, and the borderline's equation is
\begin{equation} \label{eq:BorderlineForMCUE}
|z| = \sqrt{M} ,
\end{equation}
which means that the eigenvalues f\mbox{}ill the centered circle of radius $\sqrt{M}$.


\subsection{Free Additive Unitary Dif\mbox{}fusion}
\label{s:FreeAdditiveUnitaryDiffusionCUE}

The above solution is valid for any $M \geq 2$. Now let us consider the limit
\begin{equation}
M \to \infty ,
\end{equation}
and investigate the behaviour of the eigenvalues' density and borderline's equation when $M$ grows to inf\mbox{}inity.

Obviously, this cannot be done for the model considered above, \emph{i. e.} $U_{1} + \ldots + U_{M}$, since it has no proper large--$M$ limit; the eigenvalues' density (\ref{eq:EigenvaluesDensityForMCUE}) tends to zero and the radius of the bordering circle (\ref{eq:BorderlineForMCUE}) grows to inf\mbox{}inity. Since we have the scaling relations
\begin{equation}
G_{k X} ( z , \bar{z} ) = \frac{1}{k} G_{X} \left( \frac{z}{k} , \frac{\bar{z}}{k} \right) , \qquad C_{k X} ( z , \bar{z} ) = \frac{1}{k^{2}} C_{X} \left( \frac{z}{k} , \frac{\bar{z}}{k} \right) , \qquad \rho_{k X} ( z , \bar{z} ) = \frac{1}{k^{2}} \rho_{X} \left( \frac{z}{k} , \frac{\bar{z}}{k} \right) ,
\end{equation}
for real $k$, we see that we need to choose $k = r_{\infty} / \sqrt{M}$,  where $r_{\infty}$ is an arbitrary real and positive constant, which gives, still for f\mbox{}inite $M$,
\begin{equation}
G_{r_{\infty} \frac{U_{1} + \ldots + U_{M}}{\sqrt{M}}} ( z , \bar{z} ) = \frac{\bar{z} \left( 1 - \frac{1}{M} \right)}{r_{\infty}^{2} \left( 1 - \frac{| z |^{2}}{r_{\infty}^{2} M} \right)} ,
\end{equation}
\begin{equation}
- C_{r_{\infty} \frac{U_{1} + \ldots + U_{M}}{\sqrt{M}}} ( z , \bar{z} ) = \frac{\left( 1 - \frac{1}{M^{2}} \right) \left( r_{\infty}^{2} - | z |^{2} \right)}{r_{\infty}^{4} \left( 1 - \frac{| z |^{2}}{r_{\infty}^{2} M} \right)^{2}} ,
\end{equation}
therefore the eigenvalues' density reads
\begin{equation} \label{eq:EigenvaluesDensityForInftyCUE}
\rho_{r_{\infty} \frac{U_{1} + \ldots + U_{M}}{\sqrt{M}}} ( z , \bar{z} ) = \frac{1 - \frac{1}{M}}{\pi r_{\infty}^{2} \left( 1 - \frac{| z |^{2}}{r_{\infty}^{2} M} \right)^{2}} ,
\end{equation}
and the borderline is described by
\begin{equation} \label{eq:BorderlineForInftyCUE}
|z| = r_{\infty} .
\end{equation}

Now we are ready to make the $M \to \infty$ limit. The non--holomorphic Green's function and the eigenvectors' correlator tend to
\begin{equation}
\lim_{M \to \infty} G_{r_{\infty} \frac{U_{1} + \ldots + U_{M}}{\sqrt{M}}} ( z , \bar{z} ) = \frac{\bar{z}}{r_{\infty}^{2}} ,
\end{equation}
\begin{equation}
- \lim_{M \to \infty} C_{r_{\infty} \frac{U_{1} + \ldots + U_{M}}{\sqrt{M}}} ( z , \bar{z} ) = \frac{r_{\infty}^{2} - | z |^{2}}{r_{\infty}^{4}} ,
\end{equation}
hence the limiting eigenvalues' density reads
\begin{equation}
\lim_{M \to \infty} \rho_{r_{\infty} \frac{U_{1} + \ldots + U_{M}}{\sqrt{M}}} ( z , \bar{z} ) = \frac{1}{\pi r_{\infty}^{2}} = \mathrm{const} ,
\end{equation}
whereas the borderline is independent of $M$ and remains the centered circle of radius $r_{\infty}$. In other words, the eigenvalues are uniformly distributed inside the centered circle of radius $r_{\infty}$. Note that this distribution is the same as the eigenvalues' distribution of the Girko--Ginibre model~\cite{GIRKOGINIBRE}. We can regard the described property as a certain kind of central limit theorem.


\subsection{Numerical Conf\mbox{}irmation}

We shall conf\mbox{}irm the results for a few values of $M$, namely for $M = 3 , 5 , 10$. For each of them we have drawn $100$ times $M$ unitary matrices of size $200 \times 200$ from the uniform distribution (CUE), added them, normalized through dividing by $\sqrt{M}$, and diagonalized the result. We have chosen $r_{\infty} = 1$.

The left column of the f\mbox{}igures shows, for each of the above values of $M$, positions of these $20 000$ eigenvalues on the complex plane as well as the theoretical borderline (\ref{eq:BorderlineForMCUE}). The right column presents the radial sections of the eigenvalues' density plots (again, they are circularly symmetric); the solid curves are theoretical (\ref{eq:EigenvaluesDensityForMCUE}), and the histograms are numerical.

\includegraphics[angle=-90,width=0.5\textwidth]{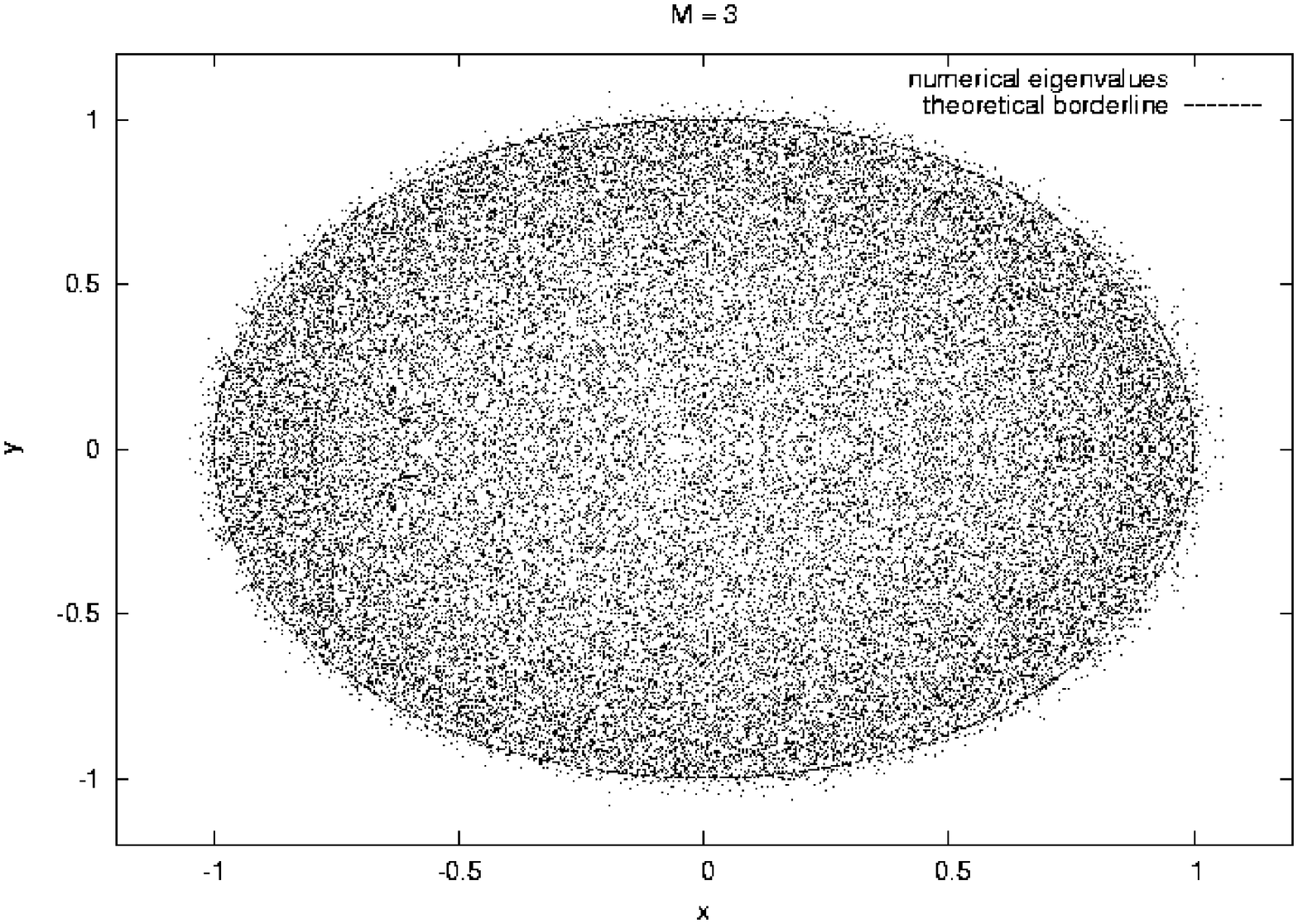}
\includegraphics[angle=-90,width=0.5\textwidth]{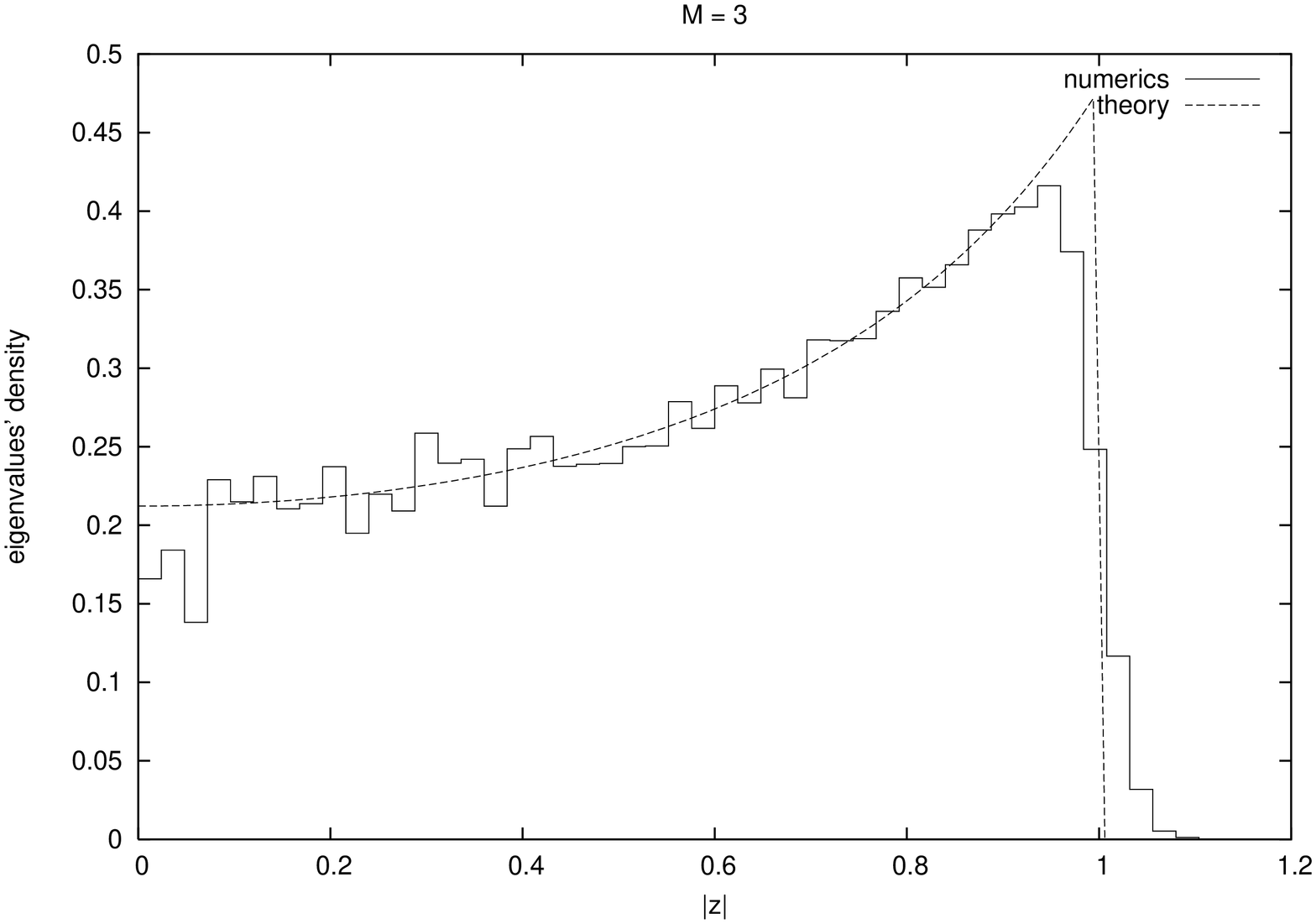}\\
\includegraphics[angle=-90,width=0.5\textwidth]{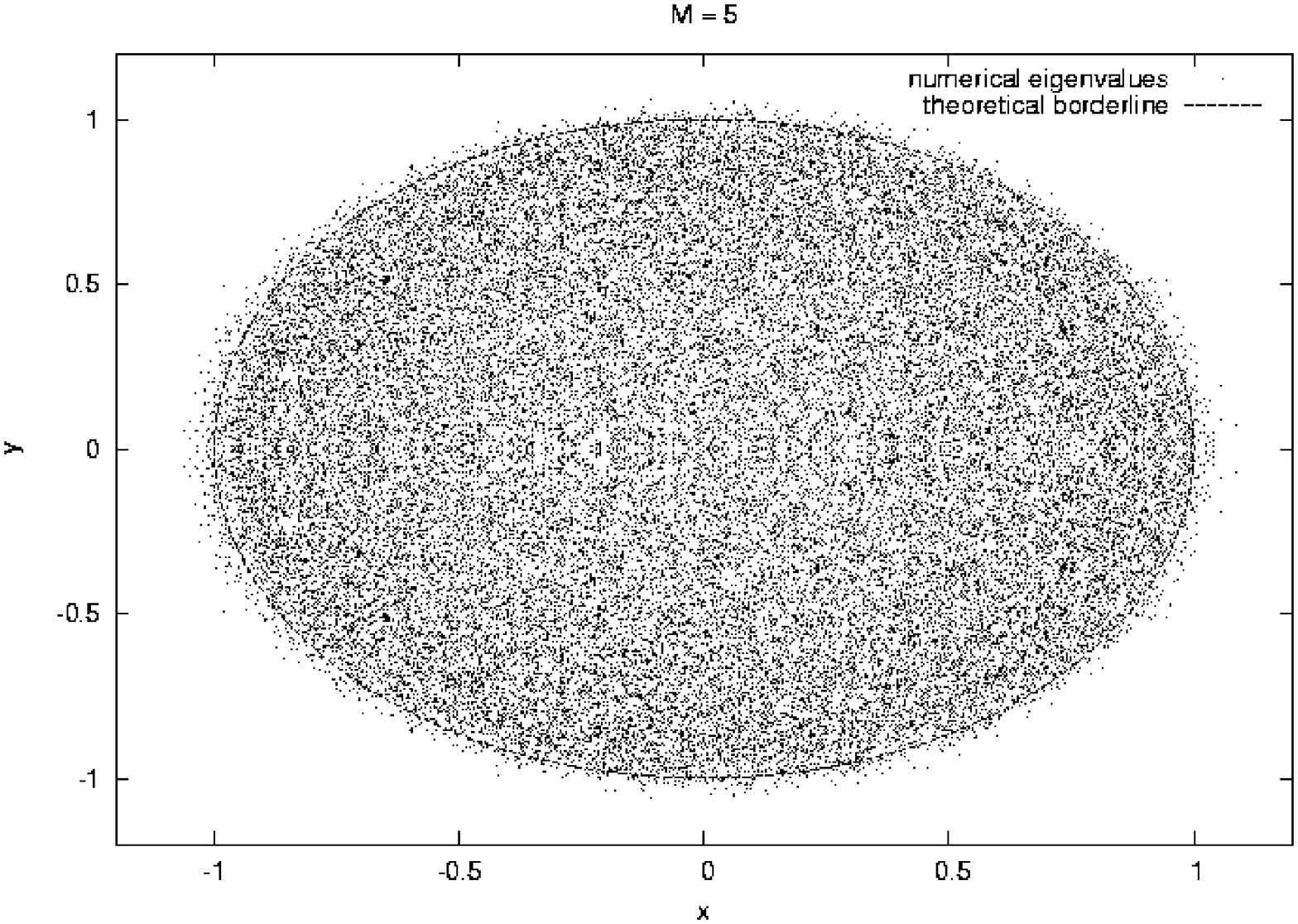}
\includegraphics[angle=-90,width=0.5\textwidth]{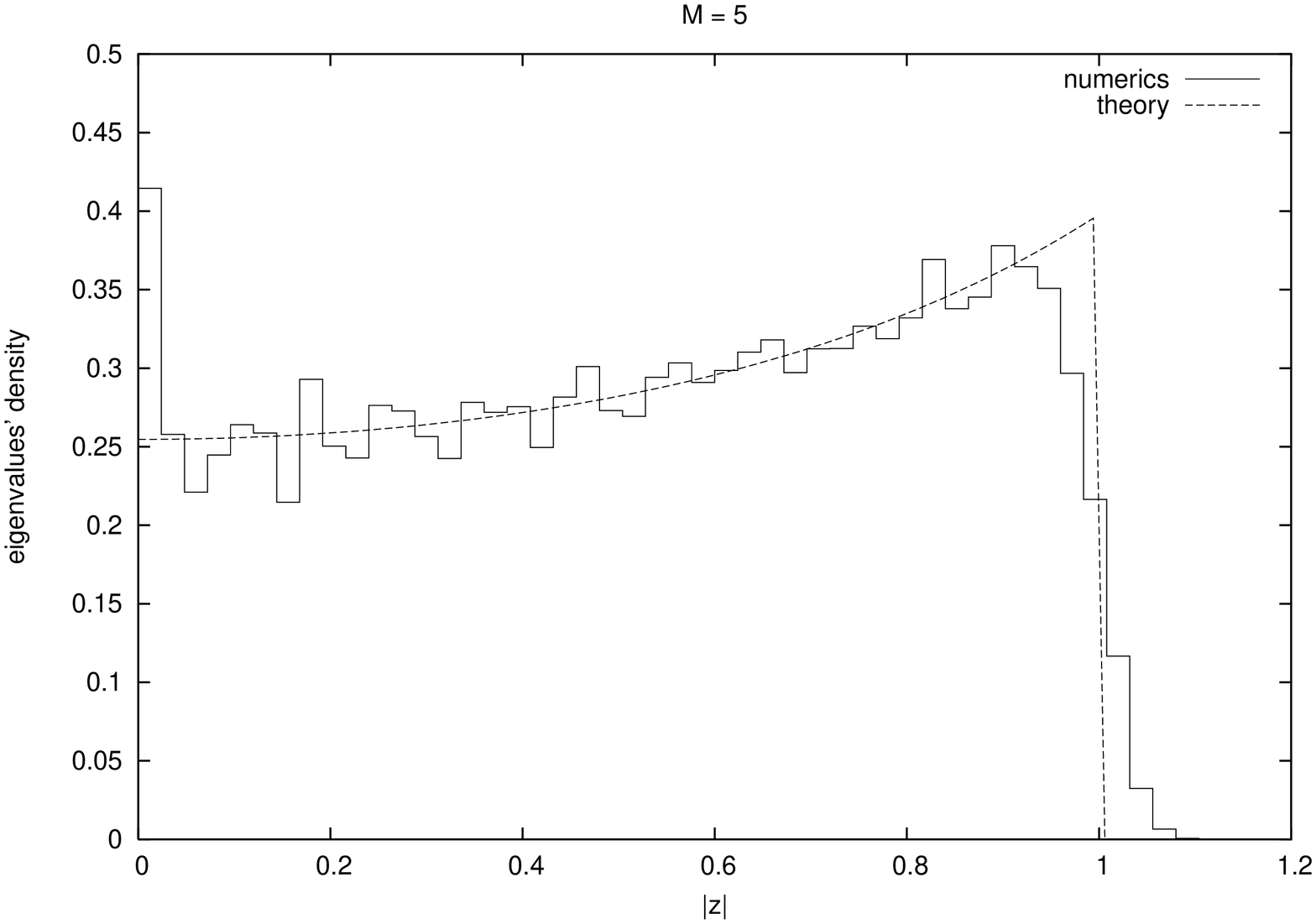}\\
\includegraphics[angle=-90,width=0.5\textwidth]{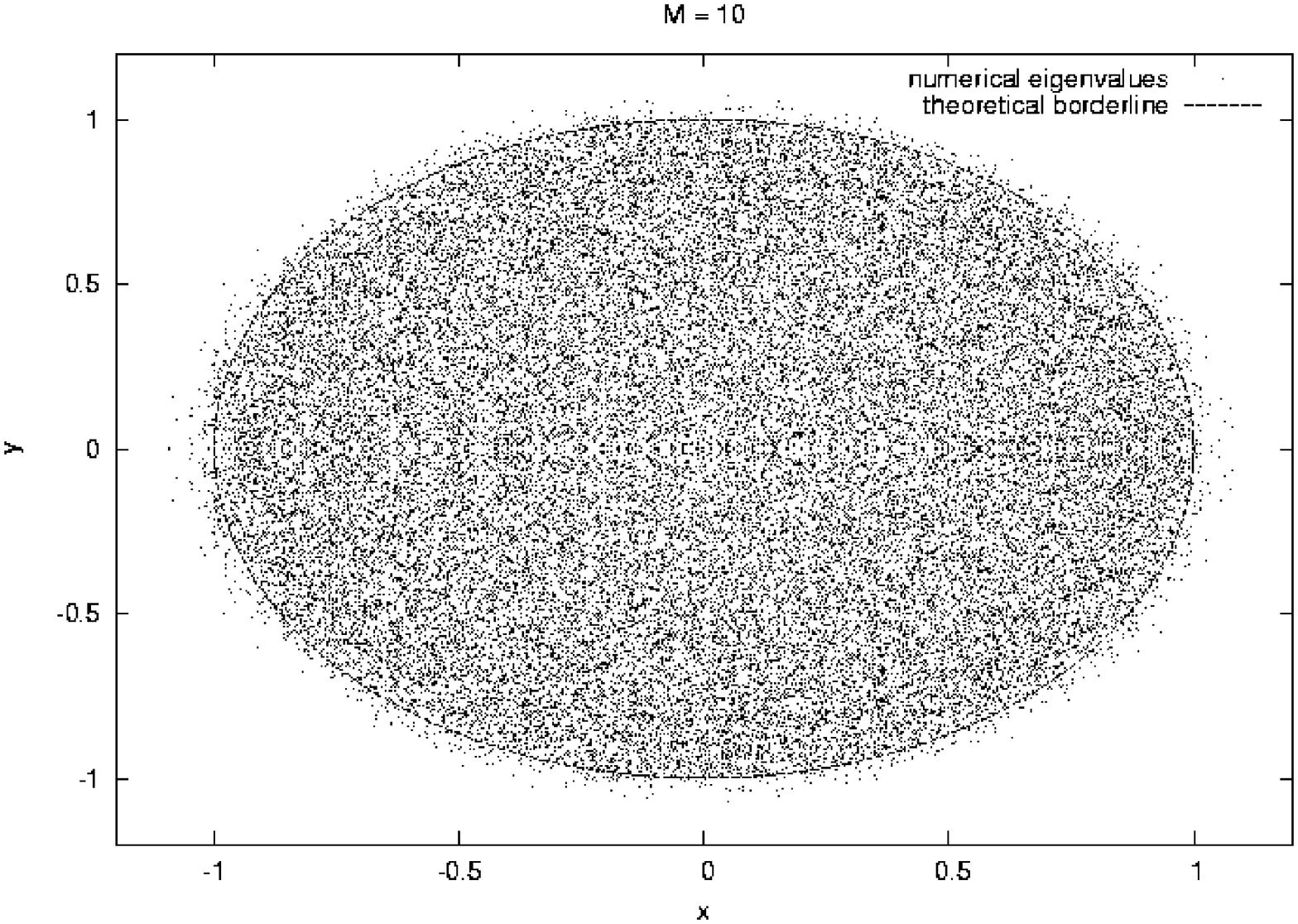}
\includegraphics[angle=-90,width=0.5\textwidth]{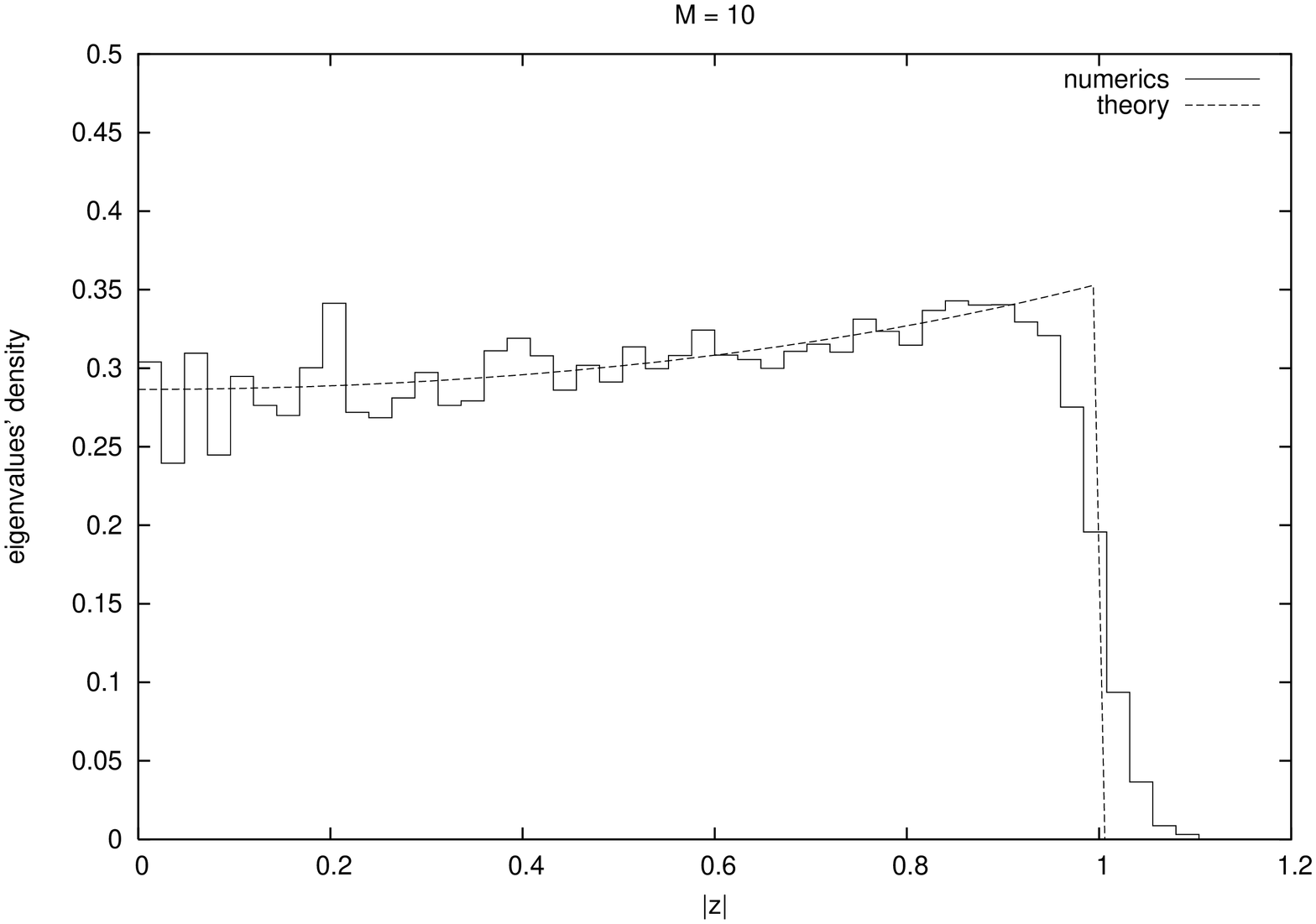}\\

Again, the complete agreement between theory and numerics is found.

We also see how the density converges into the uniform distribution on the unit circle, in agreement with (\ref{eq:EigenvaluesDensityForInftyCUE}) and (\ref{eq:BorderlineForInftyCUE}).


\section{CUE Plus GUE Model}
\label{s:CUEPlusGUEModel}


\subsection{Introduction}

Let us now solve the following model,
\begin{equation}
X = U + p H ,
\end{equation}
where $U$ is the CUE random matrix, $H$ is the GUE random matrix, and they are free, whereas $p$ is a given real constant, which may be assumed to be positive, for we shall see that $p$ appears everywhere only as $p^{2}$.

The quaternion addition law (\ref{eq:QuaternionAdditionLaw}) is now
\begin{displaymath}
\mathcal{B}_{U + p H} ( Q ) = \mathcal{B}_{U} ( Q ) + \mathcal{B}_{p H} ( Q ) - \frac{1}{Q} = \mathcal{B}_{U} ( Q ) + p^{2} Q ,
\end{displaymath}
where we have used the formula for the quaternion Blue's function for GUE and the scaling law, $\mathcal{B}_{p H} ( Q ) = p \mathcal{B}_{H} ( p Q ) = p^{2} Q + 1 / Q$, \emph{i. e.}
\begin{equation}
C = c + p^{2} a , \qquad D = d + p^{2} b .
\end{equation}

The basic equations are therefore
\begin{equation} \label{eq:BasicCUEGUE1}
z = c + p^{2} a ,
\end{equation}
\begin{equation} \label{eq:BasicCUEGUE2}
0 = d + p^{2} b .
\end{equation}
Once again, we are only to consider the generic case of $a \neq 0$, because otherwise $c = 0$, which would contradict (\ref{eq:BasicCUEGUE1}).


\subsection{Solution}

F\mbox{}irst, (\ref{eq:BasicCUEGUE1}) together with (\ref{eq:QuaternionBlueForCUE1}) show that
\begin{equation} \label{eq:3Sol1}
\alpha = z a - p^{2} a^{2} .
\end{equation}
Therefore we break $a$ into its real and imaginary parts,
\begin{equation} \label{eq:3Sol7}
a \equiv \omega + i \omega^{\prime} ,
\end{equation}
which changes (\ref{eq:3Sol1}) into two real conditions,
\begin{displaymath}
\alpha = x \omega - y \omega^{\prime} - p^{2} \omega^{2} + p^{2} \omega^{\prime 2} ,
\end{displaymath}
\begin{displaymath}
0 = x \omega^{\prime} + y \omega - 2 p^{2} \omega \omega^{\prime} .
\end{displaymath}
The second one can be used to express $\omega^{\prime}$ through $\omega$,
\begin{equation} \label{eq:3Sol8}
\omega^{\prime} = \frac{y \omega}{2 p^{2} \omega - x} ,
\end{equation}
and hence also $\alpha$ through $\omega$,
\begin{equation} \label{eq:3Sol2}
\alpha = x \omega - p^{2} \omega^{2} + p^{2} \left( \frac{y \omega}{2 p^{2} \omega - x} \right)^{2} - \frac{y^{2} \omega}{2 p^{2} \omega - x} .
\end{equation}
If we compute $\omega$, this will give also $\alpha$.

Let us note that we can express $| a |^{2}$ through $\omega$,
\begin{equation}
| a |^{2} = \omega^{2} + \omega^{\prime 2} = \omega^{2} + \left( \frac{y \omega}{2 p^{2} \omega - x} \right)^{2} .
\end{equation}
One can also check the simple identity,
\begin{equation}
\frac{\alpha}{| a |^{2}} = \frac{x}{\omega} - p^{2} ,
\end{equation}
which will soon become useful.

Moreover, (\ref{eq:BasicCUEGUE2}) states together with (\ref{eq:QuaternionBlueForCUE2}) that $\beta$ is just a constant,
\begin{equation}
\beta = p^{2} , \qquad \textrm{\emph{i. e.}} \qquad \sqrt{g^{2} - 4 \frac{\alpha^{2}}{| a |^{2}}} = p^{2} .
\end{equation}

Let us f\mbox{}irst inspect consequences of these simplif\mbox{}ications for (\ref{eq:QuaternionBlueForCUE4}), which expresses, after some manipulations, $| b |^{2}$ through $\omega$ and $\alpha$,
\begin{displaymath}
| b |^{2} p^{4} = \left( \frac{x}{\omega} - 3 p^{2} \right) \alpha + p^{2} - 1 = \ldots
\end{displaymath}
or only through $\omega$ if we exploit (\ref{eq:3Sol2}),
\begin{displaymath}
\ldots = \frac{1}{( x - 2 p \omega )^{2}} \left( x^{2} ( p^{2} - 1 + x^{2} + y^{2}) + p x ( 4 ( 1 - p^{2} ) - ( 5 + 3 p ) x^{2} - ( 1 + 3 p ) y^{2} ) \omega + \right.
\end{displaymath}
\begin{equation} \label{eq:3Sol3}
\left. + p^{2} ( 4 ( p^{2} - 1 ) + ( 8 + 15 p ) x^{2} + 3 p y^{2} ) \omega^{2} - 4 p^{3} ( 1 + 6 p ) x \omega^{3} + 12 p^{5} \omega^{4} \right) .
\end{equation}
If we compute $\omega$, this will give also $\beta$. From (\ref{eq:QuaternionBlueForCUE5}) we get
\begin{equation}
g = 2 \left( \frac{x}{\omega} - 2 p^{2} \right) \alpha + p^{2} ,
\end{equation}
and so the combination
\begin{displaymath}
g^{2} - 4 \frac{\alpha^{2}}{| a |^{2}} - \beta^{2} = 4 \left( \frac{x}{\omega} - 2 p^{2} \right)^{2} \alpha^{2} + 4 \left( \frac{x}{\omega} ( p^{2} - 1 ) + p^{2} ( 1 - 2 p^{2} ) \right) \alpha ,
\end{displaymath}
which gives due to (\ref{eq:QuaternionBlueForCUE3}) an equation for $\omega$,
\begin{equation} \label{eq:3Sol9}
\left( \frac{x}{\omega} - 2 p^{2} \right)^{2} \alpha = \frac{x}{\omega} ( 1 - p^{2} ) + p^{2} ( 2 p^{2} - 1 ) ,
\end{equation}
\emph{i. e.}, after using the explicit form of $\alpha$ (\ref{eq:3Sol2}),
\begin{equation} \label{eq:3Sol4}
( p^{2} - 1 ) x + x^{3} + x y^{2} + p^{2} ( 1 - 2 p^{2} - 5 x^{2} - y^{2} ) \omega + 8 p^{4} x \omega^{2} - 4 p^{6} \omega^{3} = 0 ,
\end{equation}
which is a desired third order (Cardano--type) equation for $\omega$. This is the solution, because we have found the equation satisf\mbox{}ied by $\omega$ and expressed $a$ and $| b |^{2}$ through $\omega$.

Let us however remind that all the calculations are done under the assumption of the generic case of $c \neq 0$ and $a \neq 0$. Let us thus investigate whether there exists the limit of
\begin{equation}
a \to 0 , \qquad \textrm{which corresponds to} \qquad \omega \to 0
\end{equation}
of our solutions. Immediately we see from (\ref{eq:3Sol3}) that $| b |^{2} p^{4}$ tends to
\begin{equation}
| b |^{2} p^{4} \to p^{2} - 1 + x^{2} + y^{2} ,
\end{equation}
hence the equation (\ref{eq:3Sol4}) for $\omega$ simplif\mbox{}ies into
\begin{displaymath}
p^{2} - 1 + x^{2} + y^{2} = 0 ,
\end{displaymath}
which means that there exists a solution of our basic set of equations which cannot be approached from the generic case, and which reads
\begin{equation} \label{eq:3Sol5}
b = 0 ,
\end{equation}
and
\begin{equation} \label{eq:3Sol6}
x^{2} + y^{2} = 1 - p^{2} .
\end{equation}
This solution exists only when (\ref{eq:3Sol6}) has solutions for $x$ and $y$, \emph{i. e.} for
\begin{equation}
p \leq 1 .
\end{equation}
Let us interpret this solution. The condition (\ref{eq:3Sol5}) of vanishing of $b$ means (\ref{eq:Borderline}) that this solution is valid on the borderline of the eigenvalues' domain. Therefore (\ref{eq:3Sol6}) says that for $p \leq 1$ at least a part of the eigenvalues' domain is a circle of radius $\sqrt{1 - p^{2}}$. We will see that this is indeed a part of the borderline, namely the internal boundary, whereas the external boundary is an ellipse. We may regard this limiting procedure as investigating the so--called \emph{holomorphic limit}~\cite{JAROSZNOWAK}, since we approach the borderline of the eigenvalues' domain.

To summarize, the solution is given by the Cardano--type equation (\ref{eq:3Sol4}) for $\omega$, which provides $a$ via (\ref{eq:3Sol7}) and (\ref{eq:3Sol8}), as well as $| b |^{2}$ via (\ref{eq:3Sol3}),
\begin{equation}
G_{U + p H} ( x , y ) = \omega + i \frac{y \omega}{2 p^{2} \omega - x} ,
\end{equation}
\begin{displaymath}
- C_{U + p H} ( x , y ) = \frac{1}{p^{4} ( x - 2 p \omega )^{2}} \left( x^{2} ( p^{2} - 1 + x^{2} + y^{2}) + p x ( 4 ( 1 - p^{2} ) - ( 5 + 3 p ) x^{2} - ( 1 + 3 p ) y^{2} ) \omega + \right.
\end{displaymath}
\begin{equation}
\left. + p^{2} ( 4 ( p^{2} - 1 ) + ( 8 + 15 p ) x^{2} + 3 p y^{2}) \omega^{2} - 4 p^{3} ( 1 + 6 p ) x \omega^{3} + 12 p^{5} \omega^{4} \right) .
\end{equation}
The eigenvalues' density follows thus immediately,
\begin{displaymath}
\rho_{U + p H} ( x , y ) = \frac{1}{2 \pi} \left( \partial_{x} + i \partial_{y} \right) \left( \omega + i \omega^{\prime} \right) = \frac{1}{2 \pi}\left( \partial_{x} \omega - \partial_{y} \omega^{\prime} \right) = \ldots
\end{displaymath}
where one can check that $\partial_{x} \omega^{\prime} + \partial_{y} \omega = 0$, so that the imaginary part vanishes, as expected. Exploiting (\ref{eq:3Sol2}) we have further
\begin{displaymath}
\ldots = \frac{1}{2 \pi} \left( \partial_{x} \omega + \partial_{y} \omega \frac{x y}{( 2 p^{2} \omega - x )^{2}} - \frac{\omega}{2 p^{2} \omega - x}\right) = \ldots
\end{displaymath}
Dif\mbox{}ferentiating now (\ref{eq:3Sol4}) with respect to $x$ and $y$ we get
\begin{displaymath}
\partial_{x} \omega = \frac{1 - p^{2} - 3 x^{2} - y^{2} + 10 p^{2} x \omega - 8 p^{4} \omega^{2}}{p^{2} ( 1 - 2 p^{2} - 5 x^{2} - y^{2} ) + 16 p^{4} x \omega - 12 p^{6} \omega^{2}} ,
\end{displaymath}
\begin{displaymath}
\partial_{y} \omega = \frac{2 y ( p^{2} \omega - x )}{p^{2} ( 1 - 2 p^{2} - 5 x^{2} - y^{2} ) + 16 p^{4} x \omega - 12 p^{6} \omega^{2}} ,
\end{displaymath}
therefore f\mbox{}inally
\begin{displaymath}
\rho_{U + p H} ( x , y ) =
\end{displaymath}
\begin{displaymath}
= \left( x^{2} ( 1 - p^{2} ) - 3 x^{4} - 3 x^{2} y^{2} + p^{2} x ( - 3 + 2 p^{2} + 17 x^{2} + 5 y^{2} ) \omega + 2 p^{4} ( 1 - 17 x^{2} - y^{2} ) \omega^{2} + 28 p^{6} x \omega^{3} - 8 p^{8} \omega^{4} \right) /
\end{displaymath}
\begin{displaymath}
/ 2 \pi \left( - p^{2} ( x^{2} ( 2 p^{2} - 1 ) + 5 x^{4} + x^{2} y^{2} ) - 4 p^{4} x ( 1 - 2 p^{2} - 9 x^{2} - y^{2} ) \omega - \right.
\end{displaymath}
\begin{equation} \label{eq:3Sol12}
\left. - 4 p^{6} ( - 1 + 2 p^{2} + 24 x^{2} + y^{2} ) \omega^{2} + 112 p^{8} x \omega^{3} - 48 p^{10} \omega^{4} \right) ,
\end{equation}
where $\omega$ is given by (\ref{eq:3Sol4}). This is the desired eigenvalues' density in a parametric form.

We need also the borderline's equation, which is $| b |^{2} = 0$, \emph{i. e.} from (\ref{eq:3Sol3}),
\begin{displaymath}
\alpha = \frac{1 - p^{2}}{\frac{x}{\omega} - 3 p^{2}} ,
\end{displaymath}
which put into (\ref{eq:3Sol9}) gives a surprisingly simple solution
\begin{displaymath}
\omega = \frac{x}{1 + 2 p^{2}} ;
\end{displaymath}
this value of $\omega$ (valid when on the borderline) should be substituted to (\ref{eq:3Sol3}) to get the borderline's equation,
\begin{equation} \label{eq:3Sol10}
\frac{1 + p^{2}}{( 1 + 2 p^{2} )^{2}} x^{2} + ( 1 + p^{2} )y^{2} = 1 ,
\end{equation}
which is an ellipse. If $p \leq 1$, this is a part of the borderline, because we know that the borderline can be reached also via the $a \to 0$ limit, which has given the circle
\begin{equation} \label{eq:3Sol11}
x^{2} + y^{2} = 1 - p^{2} .
\end{equation}
To summarize, if $p < 1$, the eigenvalues of the $U + p H$ model are placed within the domain with two boundaries, the external one, which is the ellipse (\ref{eq:3Sol10}) and the internal one which is the circle (\ref{eq:3Sol11}); for $p = 1$ the internal circle reduces to the point $( 0 , 0 )$, so to vanish for $p > 1$, and the eigenvalues f\mbox{}ill the whole ellipse (\ref{eq:3Sol10}). We have therefore a phenomenon at $p = 1$ which may be called a \emph{topological phase transition}.


\subsection{Numerical Conf\mbox{}irmation}

The results are conf\mbox{}irmed for a few values of $p$, namely for $p = 0.5 , 0.75 , 1 , 2$. For each of them we have drawn $50$ unitary matrices of size $100 \times 100$ from the uniform distribution (CUE), as well as $50$ Hermitian matrices of the same size from the Gaussian distribution (GUE), added them with appropriate $p$, and diagonalized the sum.

The f\mbox{}irst four f\mbox{}igures show, for each of the above values of $p$, positions of these $5 000$ eigenvalues on the complex plane as well as the theoretical borderlines (\ref{eq:3Sol10}) and (for $p \leq 1$) (\ref{eq:3Sol11}). We see in particular that the internal circle decreases as $p$ increases, and vanishes completely for $p = 1$ (topological phase transition), so that for $p > 1$ there remains only the external ellipse.

\includegraphics[angle=-90,width=0.5\textwidth]{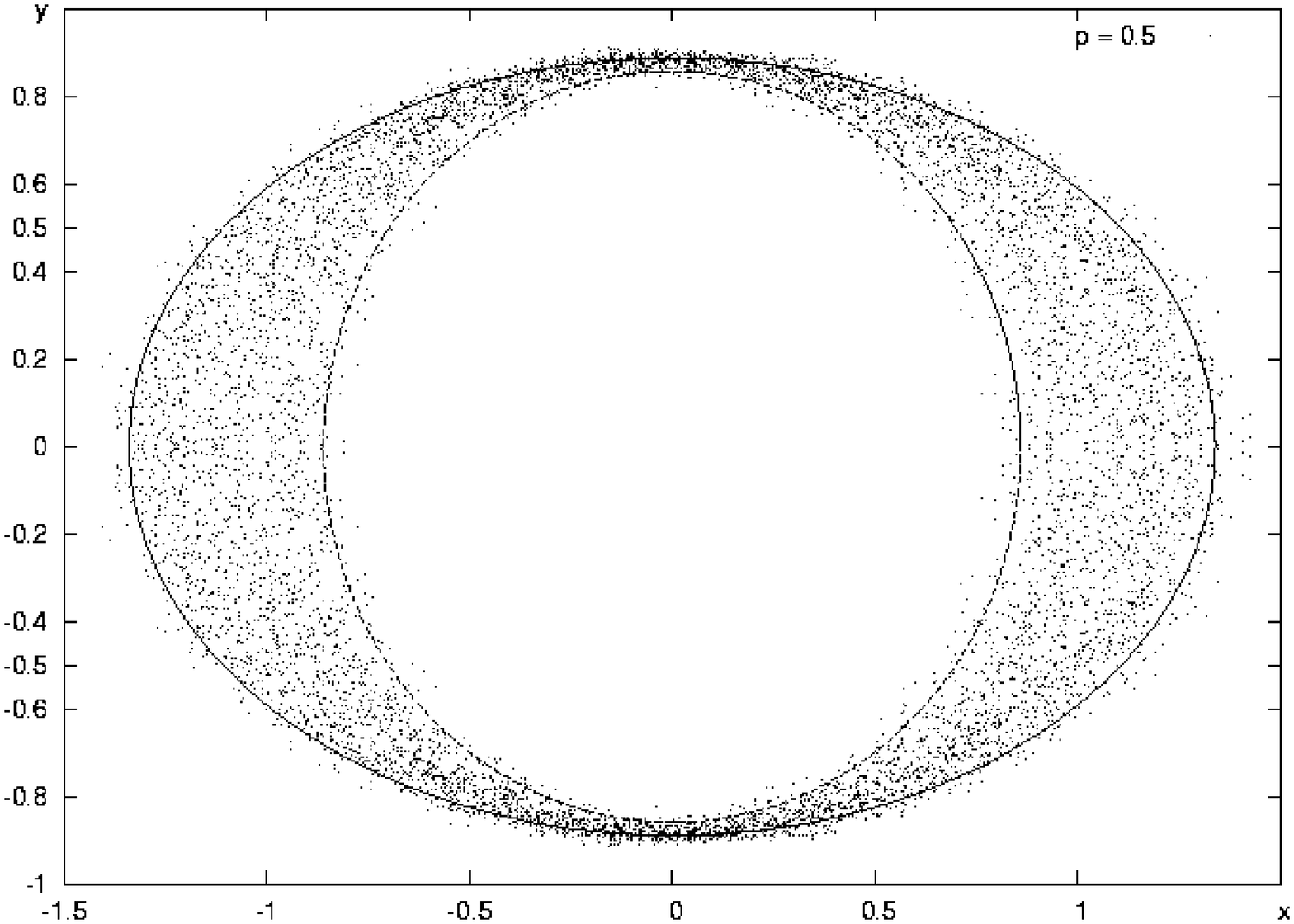}
\includegraphics[angle=-90,width=0.5\textwidth]{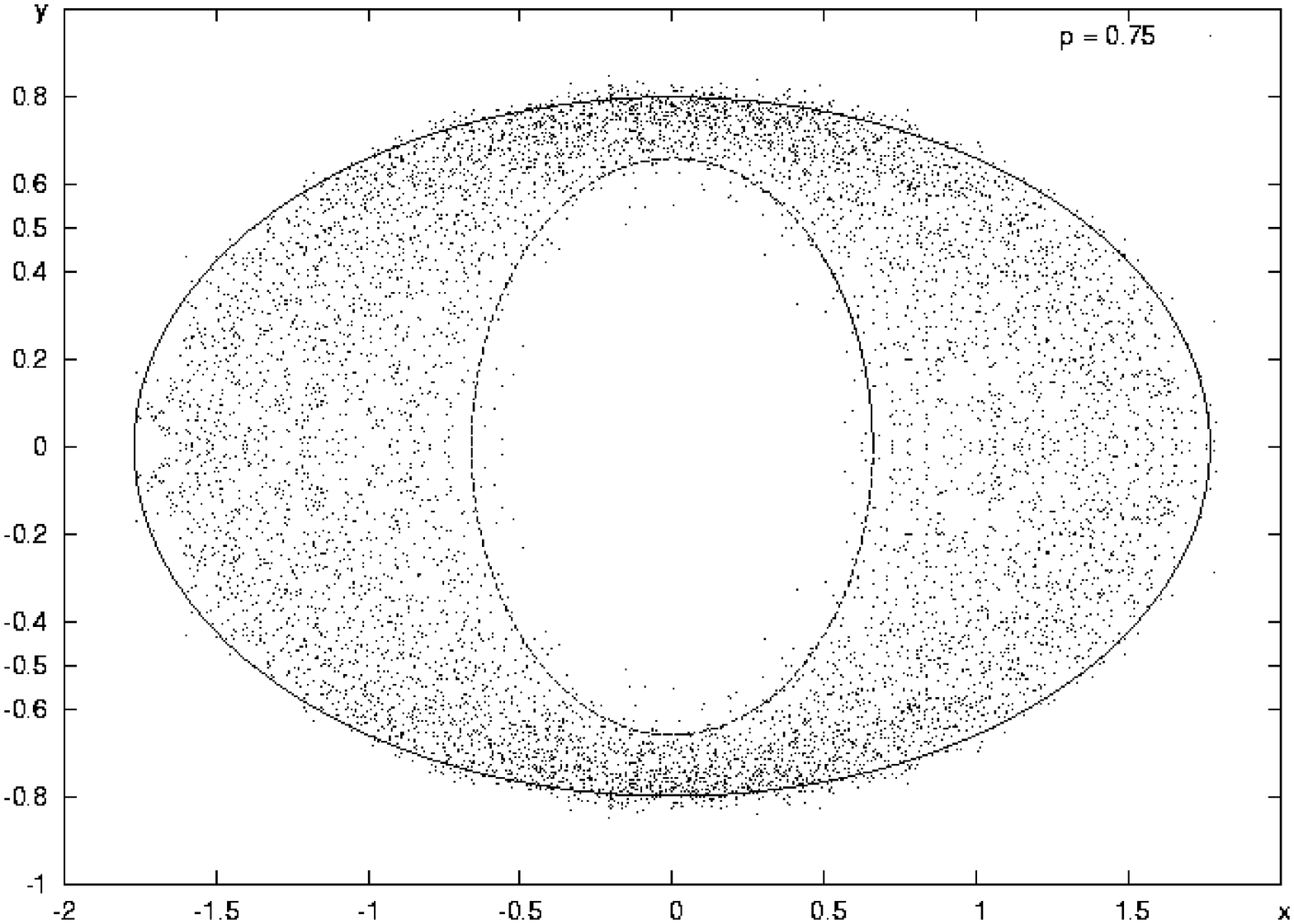}\\
\includegraphics[angle=-90,width=0.5\textwidth]{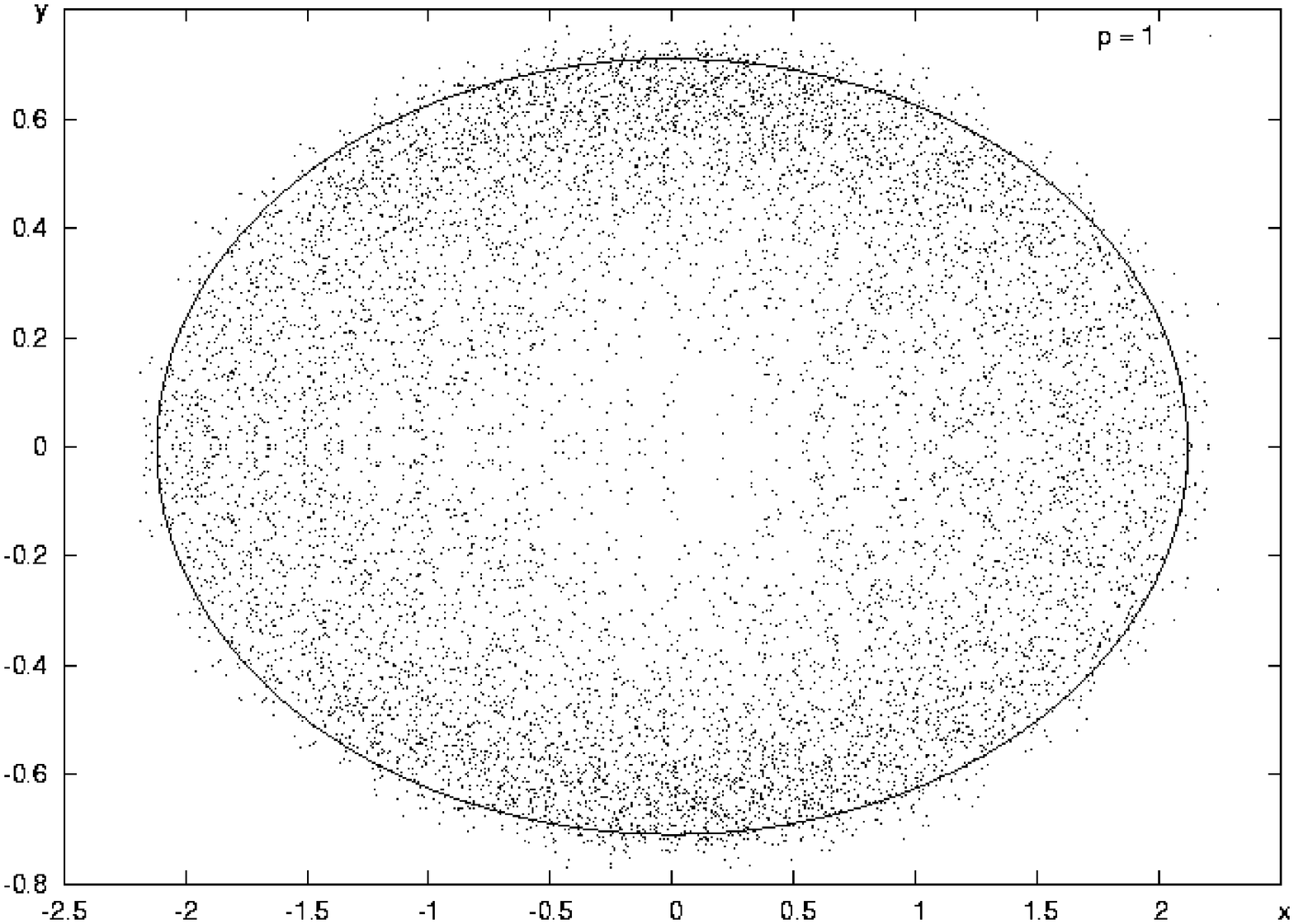}
\includegraphics[angle=-90,width=0.5\textwidth]{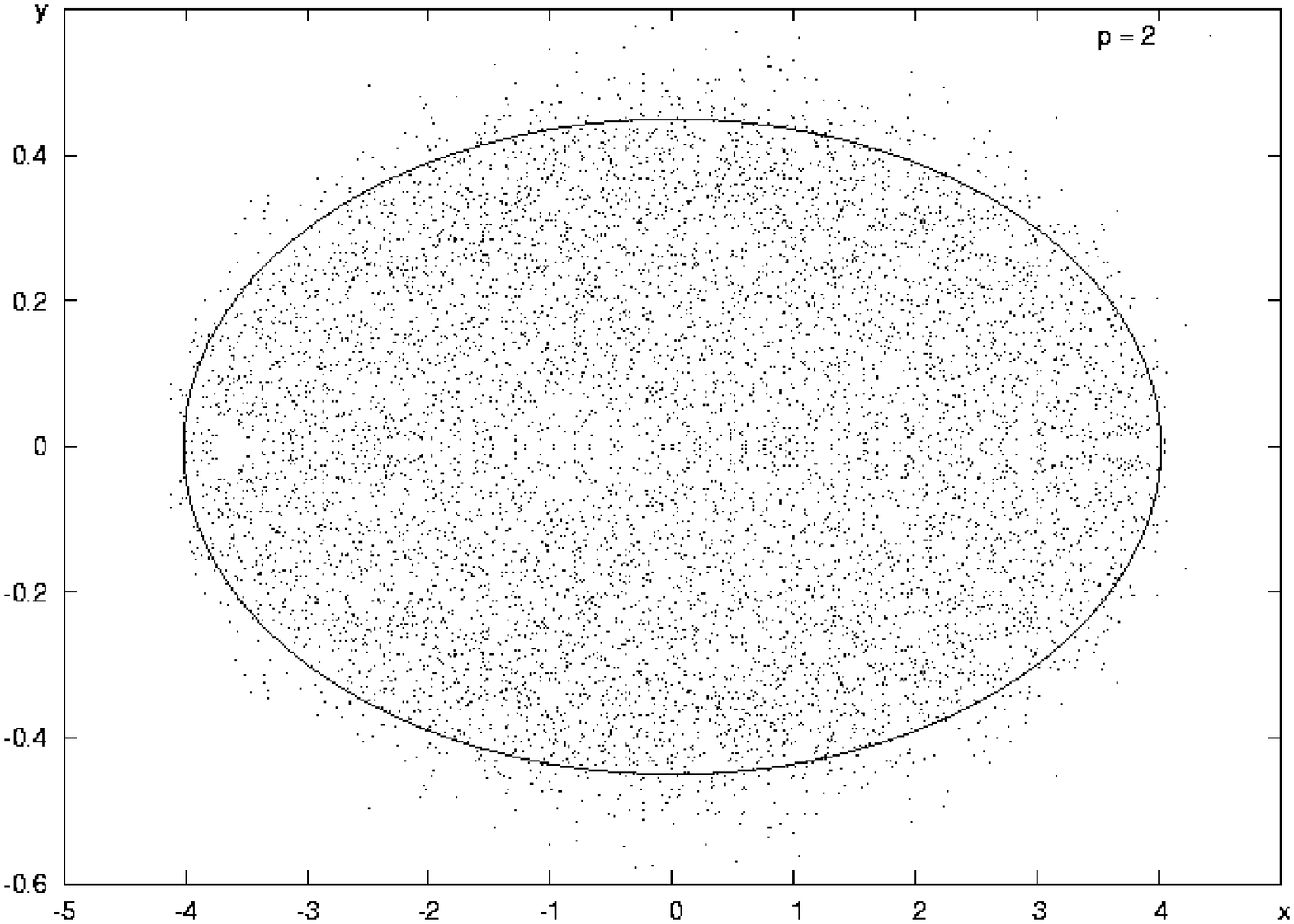}\\

Experimental points lie exactly inside the theoretically predicted domains.

To make our f\mbox{}igures even more expressive when investingating the eigenvalues' density, we have increases the statistic of matrices, namely we have drawn $100$ sets of $200 \times 200$ CUEs and GUEs, added them appropriately, and diagonalized the sum. We have repeated it for $p = 0.5 , 0.75 , 1 , 2$.

The left column of f\mbox{}igures below contains numerical histograms in three dimensions of the density, based on $20 000$ experimental points each time, whereas on the right we see the theoretical prediction (\ref{eq:3Sol12}).

\includegraphics[angle=-90,width=0.5\textwidth]{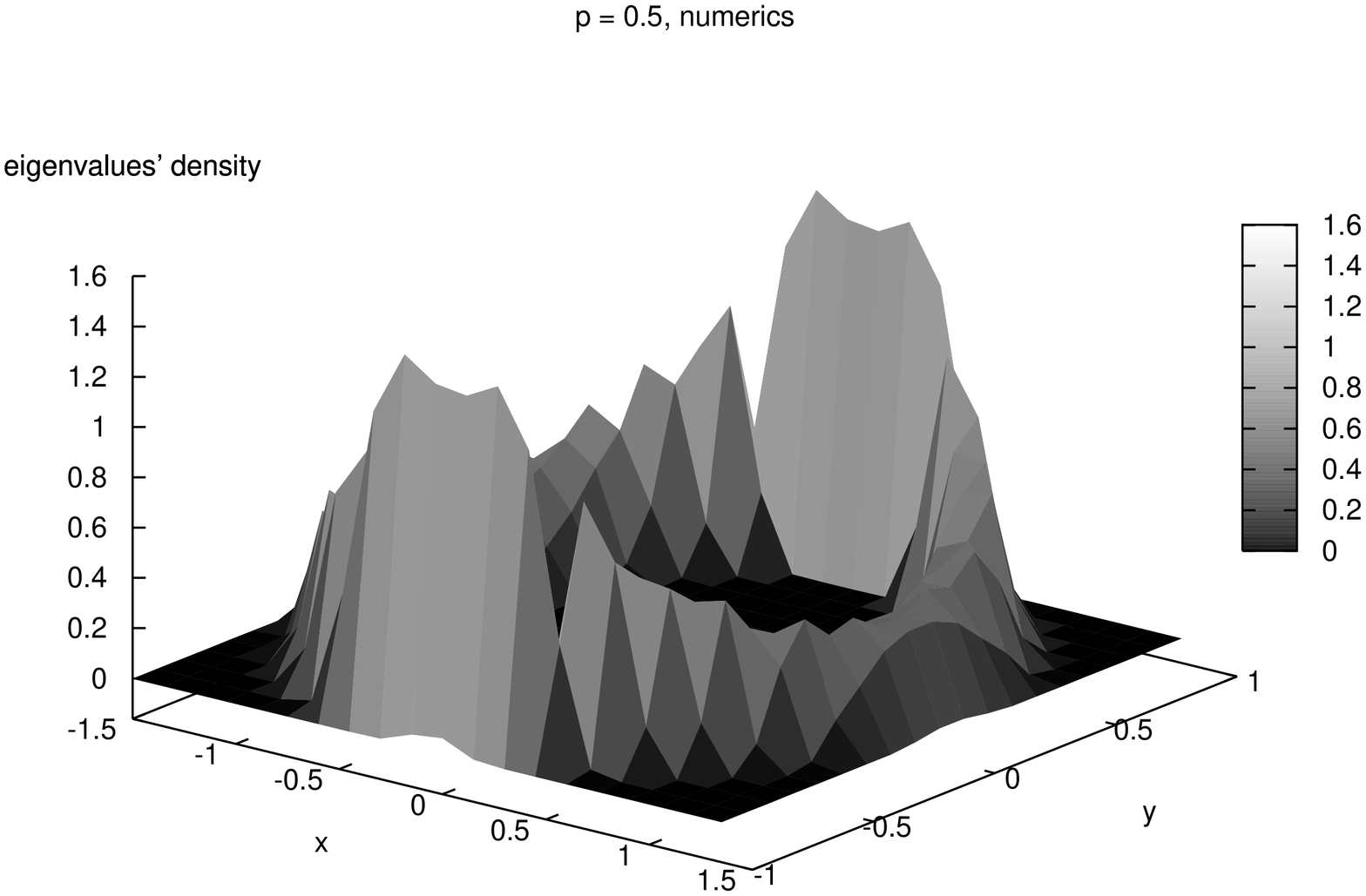}
\includegraphics[angle=-90,width=0.5\textwidth]{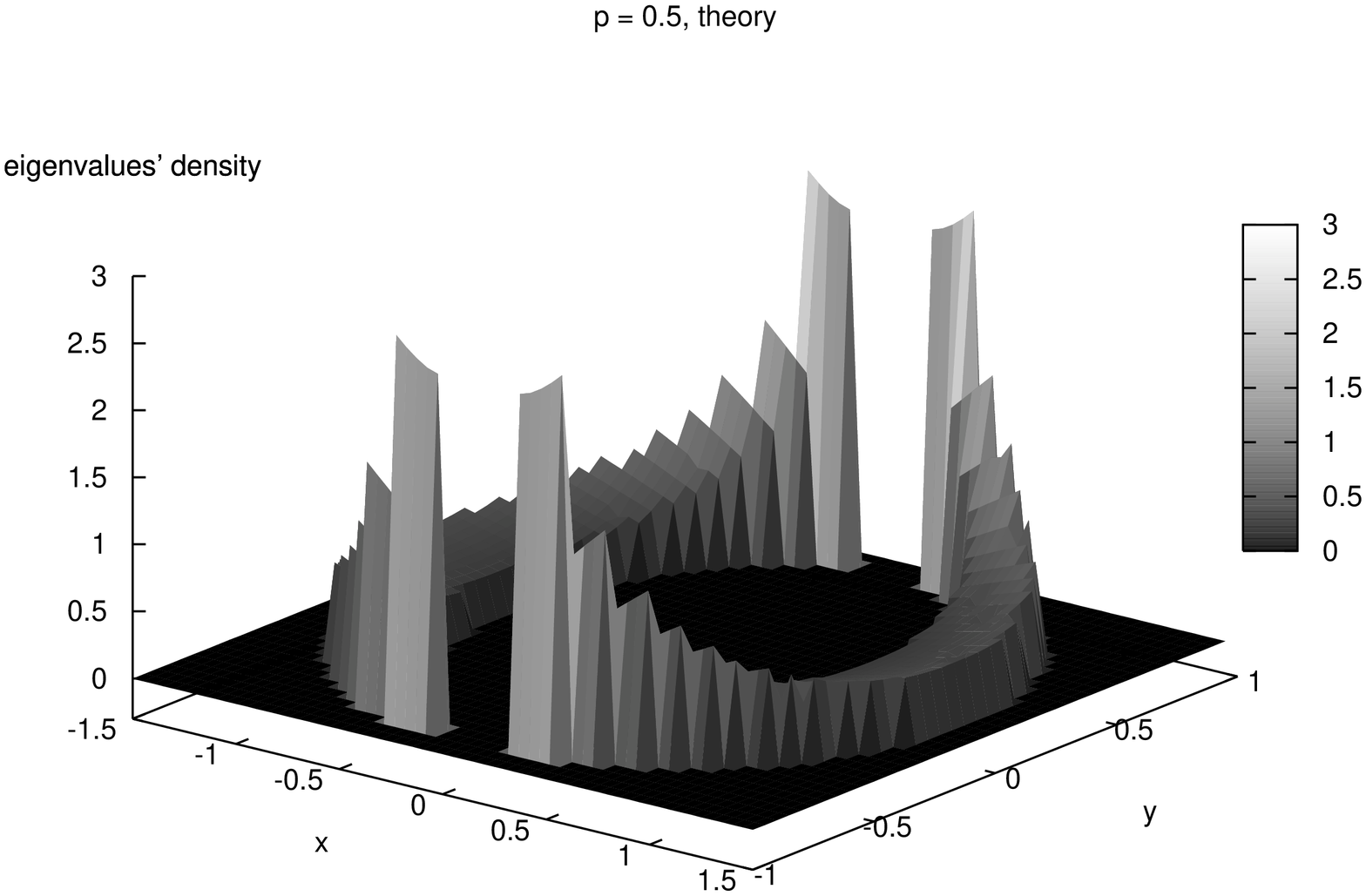}\\
\includegraphics[angle=-90,width=0.5\textwidth]{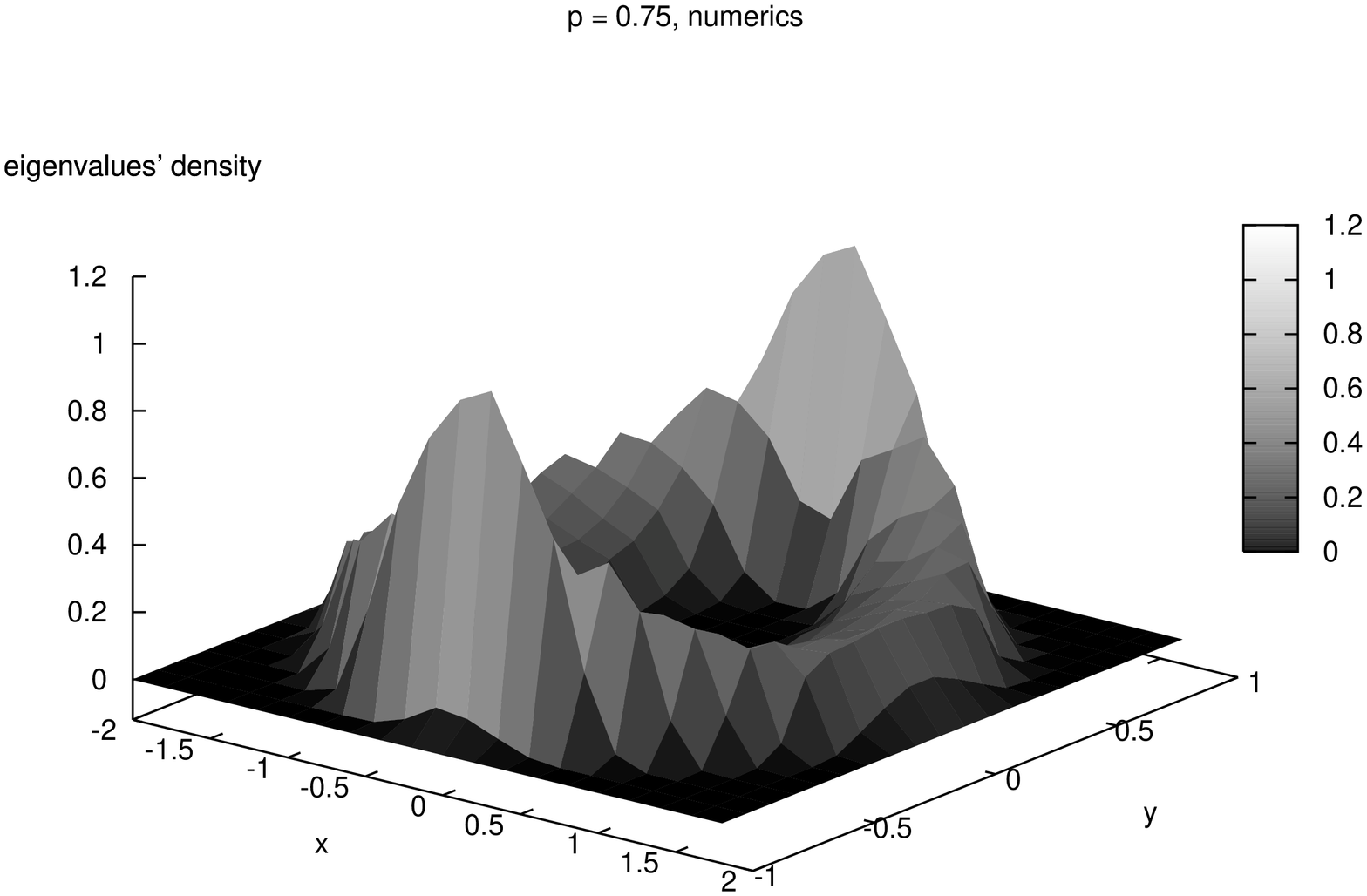}
\includegraphics[angle=-90,width=0.5\textwidth]{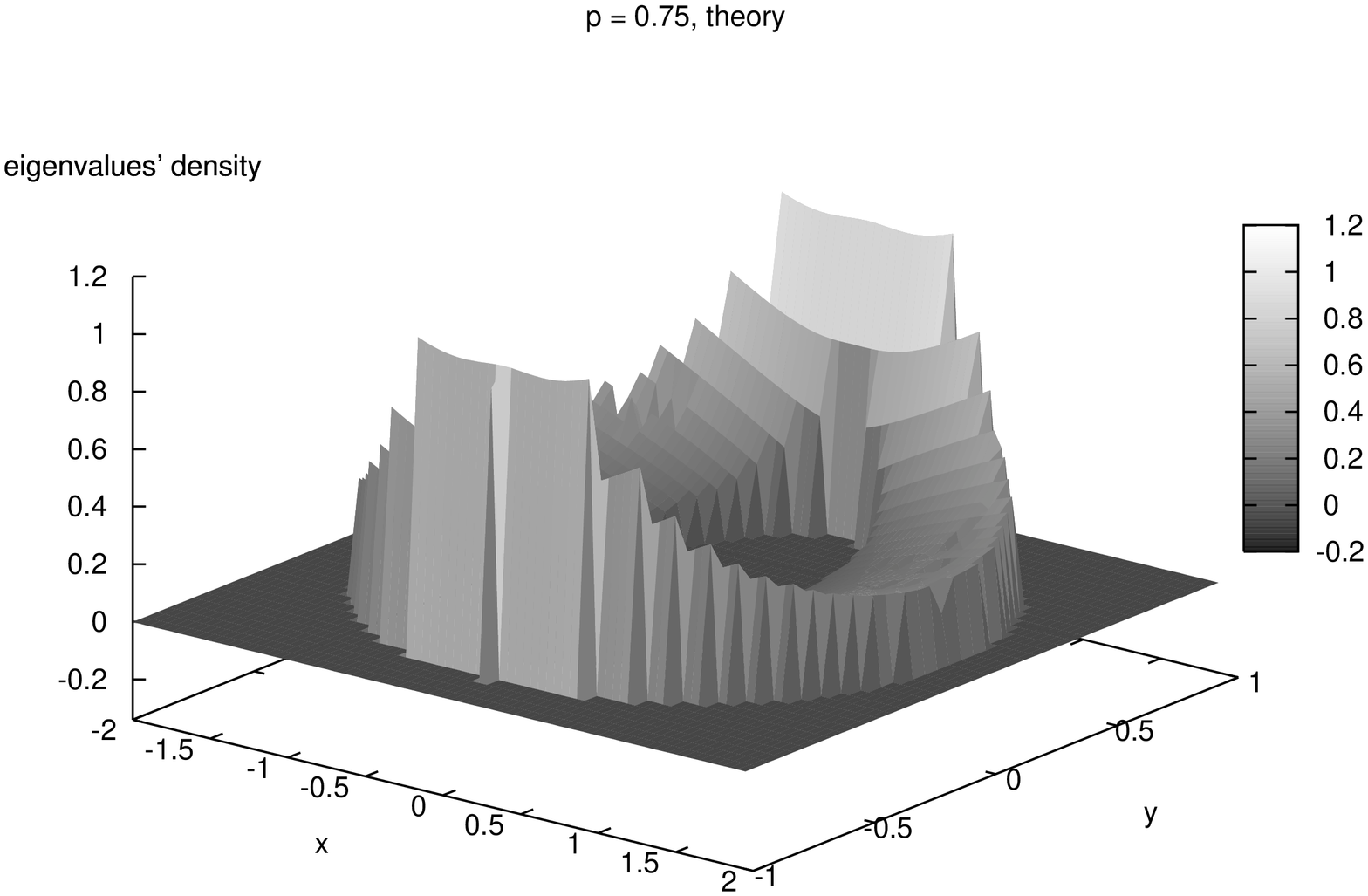}\\
\includegraphics[angle=-90,width=0.5\textwidth]{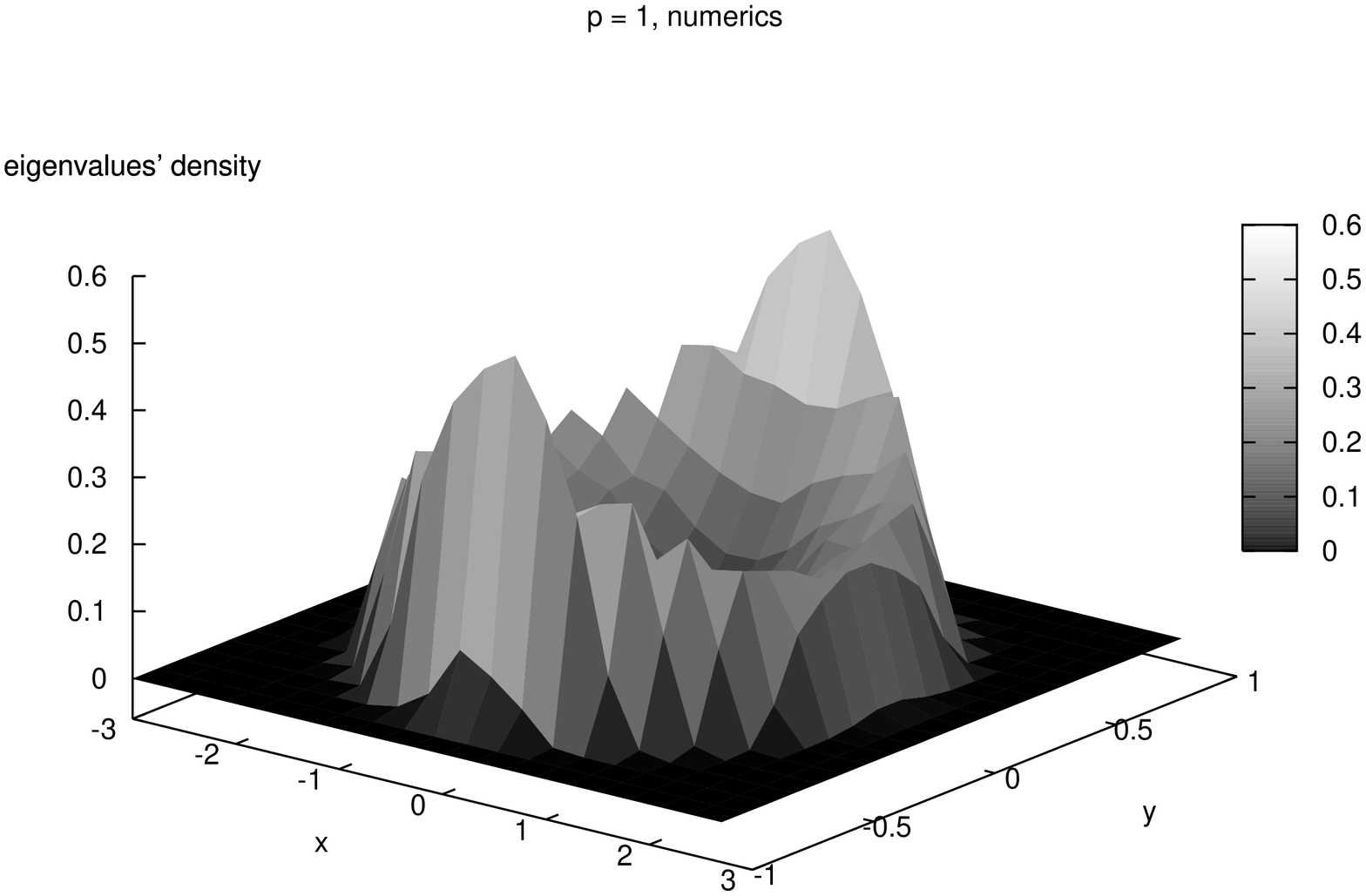}
\includegraphics[angle=-90,width=0.5\textwidth]{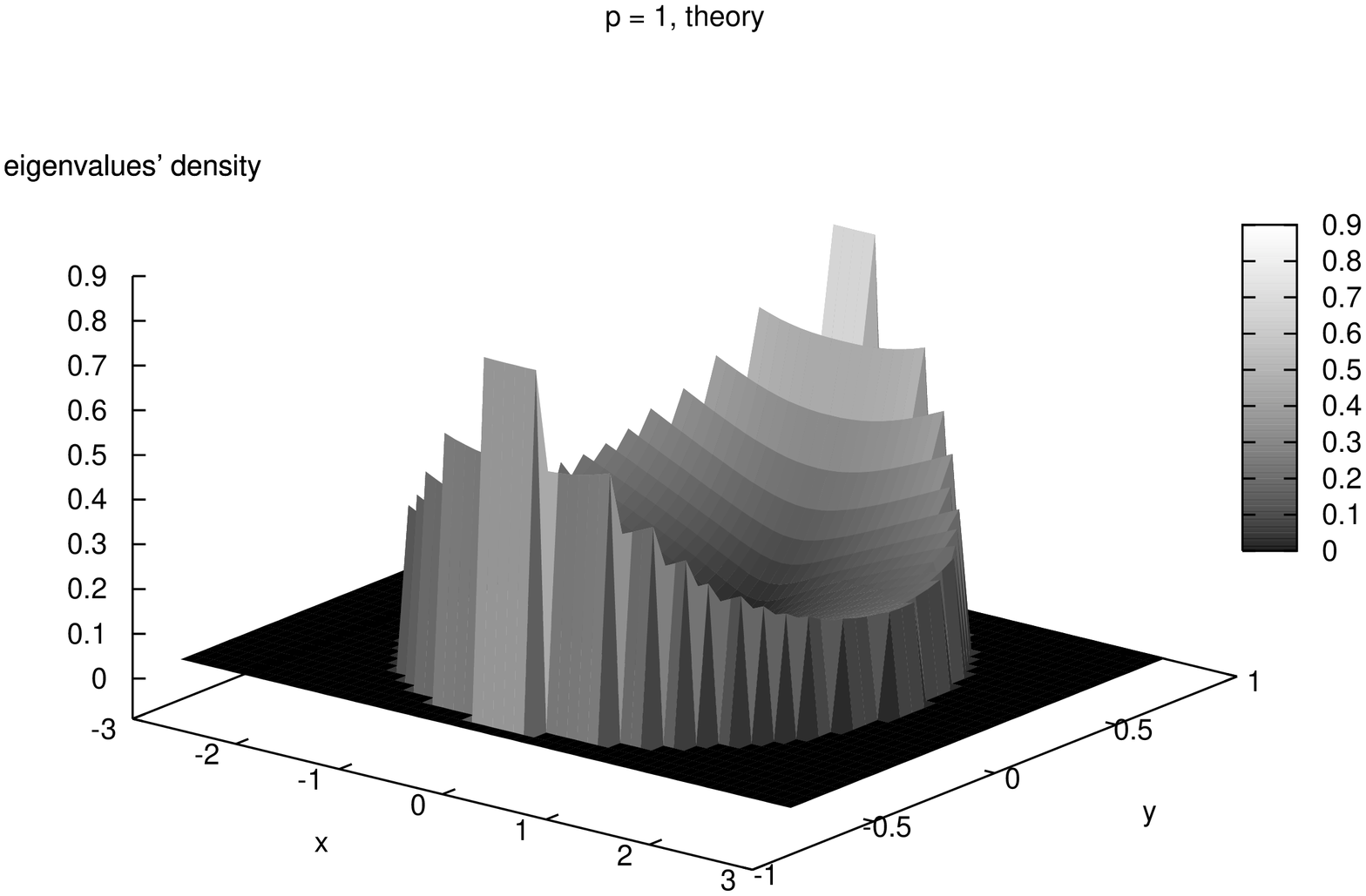}\\
\includegraphics[angle=-90,width=0.5\textwidth]{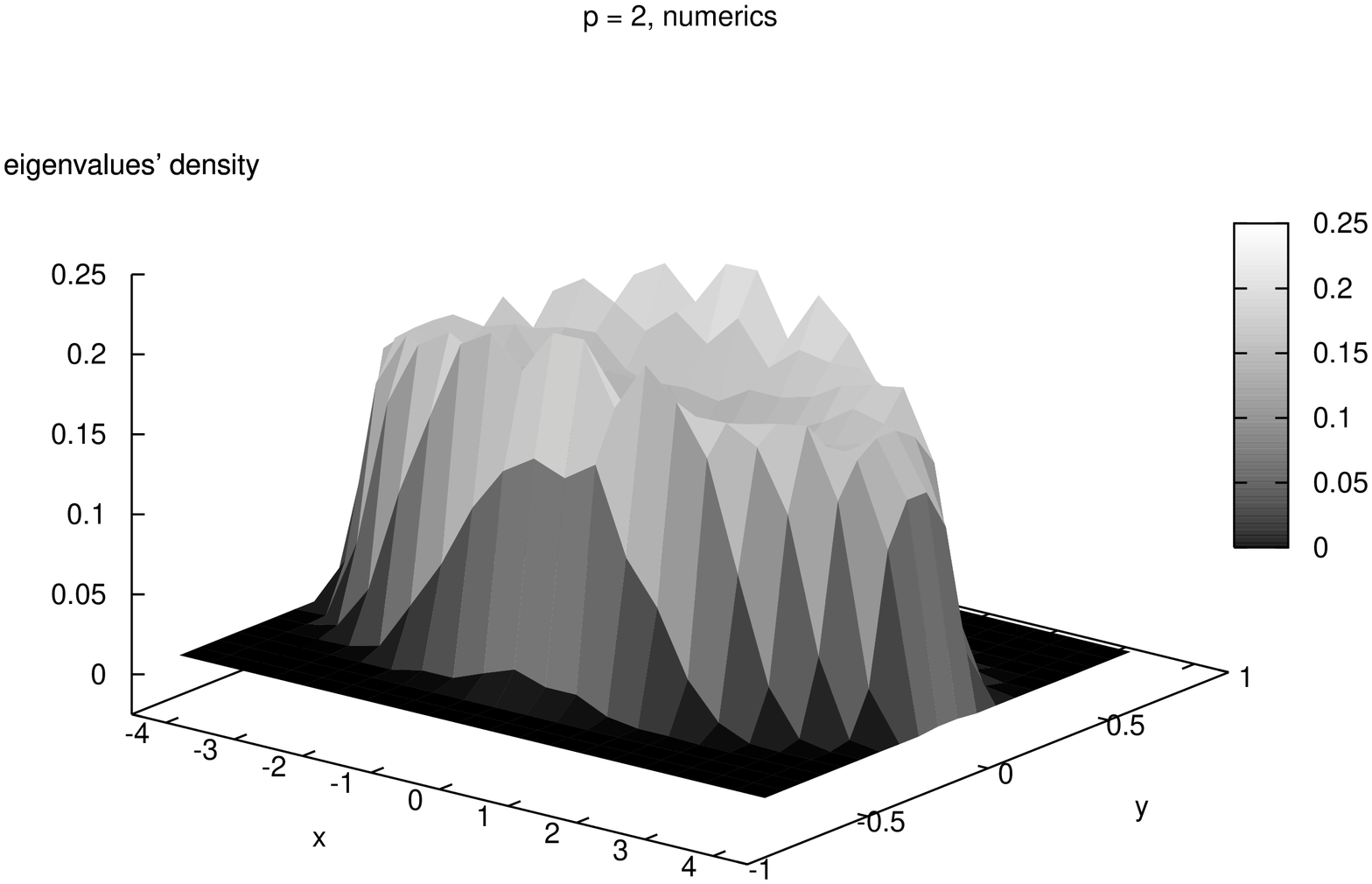}
\includegraphics[angle=-90,width=0.5\textwidth]{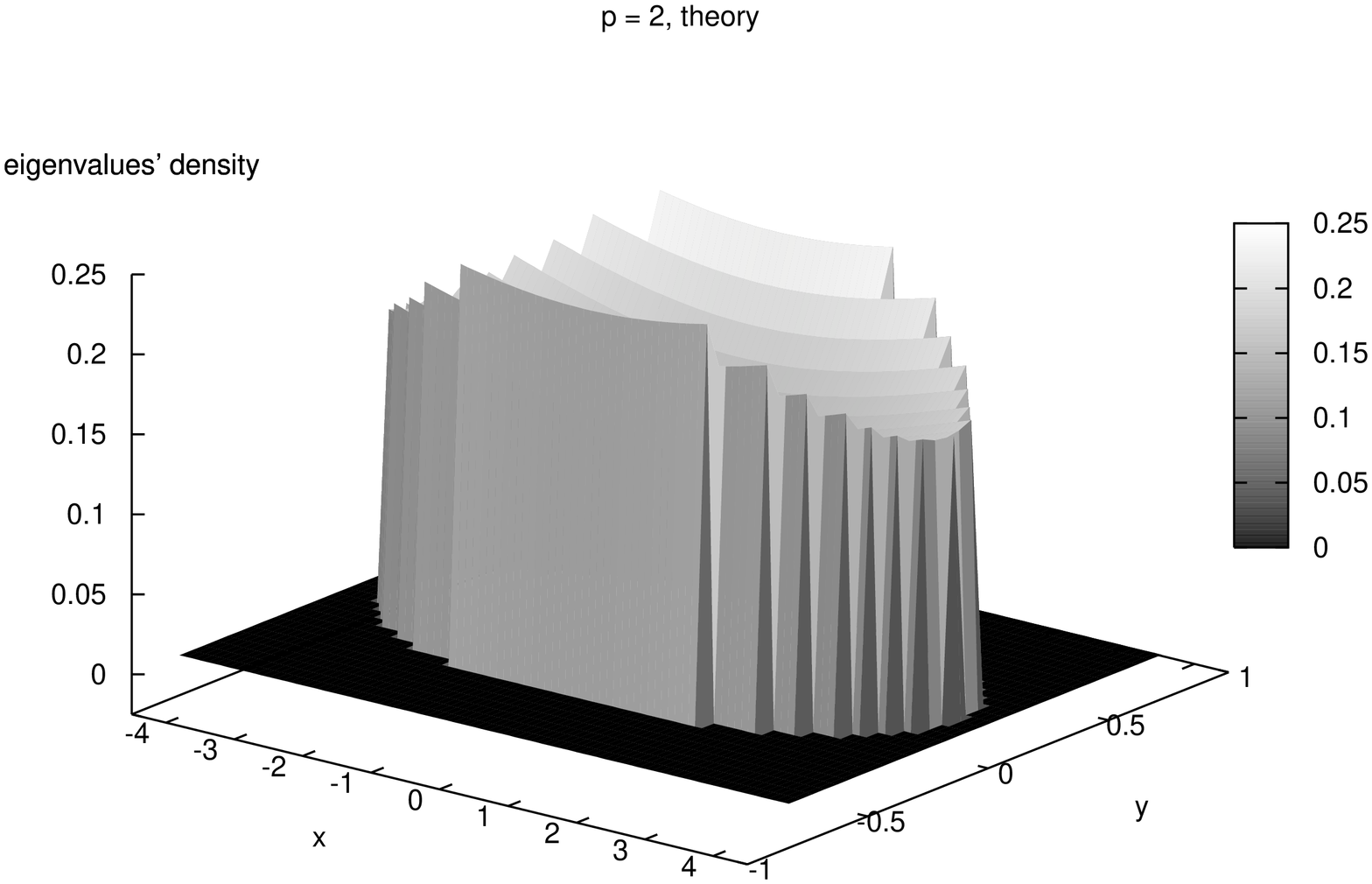}\\

We recognize the plots to be similar. It is striking that such a complicated expression as (\ref{eq:3Sol12}) is conf\mbox{}irmed numerically in such a good way. Slight dif\mbox{}ferences between the detailed shapes of the respective f\mbox{}igures are due to numerical obstacles in plotting three dimensional numerical histograms, as well as solving numerically the equation (\ref{eq:3Sol4}), choosing the proper solution, and inserting it into the expression (\ref{eq:3Sol12}) for the eigenvalues' density. (In particular, the two holes in the theoretical f\mbox{}igure for $p = 0.5$ are artif\mbox{}icial and mean that the plot is relatively thin there.)


\section{Summary and Prospects}
\label{s:SummaryAndProspects}

In this paper we have concentrated on explicit presentation of mathematical aspects of a particular application of non--Hermitian free random variables calculus to models which have the form of sums involving free unitary random matrices. We have derived general identities and solved three models as examples.

The main aim of the paper is to convince the reader that the quaternion method of\mbox{}fers a strikingly simple machinery to deal with certain instances of non--Hermitian models, reducing their solutions to elementary algebra. Sums of large free unitary matrices, or unitary and Hermitian, look at the f\mbox{}irst sight very complicated, but the quaternion technique provides a way to deal with them which involves only basic algebra.

One question that can be posed deals with possible generalizations of the method to ensembles other than Hermitian and unitary used as ``building blocks'' of non--Hermitian models. This will depend on possible evaluation methods of (\ref{eq:QuaternionGreenExplicit}); in particular, this is doable when $X$ satisf\mbox{}ies (\ref{eq:Rational}).

One can also ask about some physical applications. The problem is that unitary matrices are not usually being added, but rather multiplied. The authors cannot see any applications at the moment. However, we see the solutions of sec.~\ref{s:CUEPlusCUEModel}, \ref{s:CUEPlusldotsPlusCUEModel} and \ref{s:CUEPlusGUEModel} as at least mathematically intresting, because we can precisely trace the way the unitary noise acts. For example, for the CUE plus GUE model, we would expect that the eigenvalues of the CUE matrix, lying on the unit circle, get smeared somehow when we impose the GUE noise; and this is exactly conf\mbox{}irmed by our solution, which shows how the initial circle is smeared into the area between an internal circle and external ellipse. The three solutions we presented are thus quite intuitive, however we emphasize that the mathematical structure of the models is nevertheless involved, and this is the quaternion technique that allowed to tackle them.

Another point may be to consider the general free additive unitary dif\mbox{}fusion, which means the same inf\mbox{}inite sum,
\begin{equation}
\frac{U_{1} + \ldots + U_{M}}{\sqrt{M}} ,
\end{equation}
as in subsec.~\ref{s:FreeAdditiveUnitaryDiffusionCUE}, but for identically distributed arbitrary free unitary random matrices $U_{i}$, where eventually we are interested in the limit $M \to \infty$. This requires taking the equations (\ref{eq:BasicMCUE1}) and (\ref{eq:BasicMCUE2}), together with general (\ref{eq:QuaternionBlueForU}), and plugging into them the expansions
\begin{equation}
a = a_{0} + \frac{1}{\sqrt{M}} a_{1} + \frac{1}{M} a_{2} + \ldots , \qquad b = b_{0} + \frac{1}{\sqrt{M}} b_{1} + \frac{1}{M} b_{2} + \ldots ,
\end{equation}
where we also write the standard Green's function for $U$ as moment expansion,
\begin{equation}
G_{U} ( z ) = \frac{1}{z} + \frac{m_{U , 1}}{z^{2}} + \frac{m_{U , 2}}{z^{3}} + \ldots .
\end{equation}
Now comparison of appropriate terms in large--$M$ expansions of the formuale will give some equations for $a_{0}$, $a_{1}$, \ldots, and $b_{0}$, $b_{1}$, \ldots. Unfortunately, to our present knowledge, these equations are quite complicated and it may require a lot to extract the solution. This result will be an additive analog to the computation~\cite{JANIKWIECZOREK} of the free multiplicative unitary dif\mbox{}fusion.


\section*{Acknowledgements}

The authors are grateful to Z. Burda, R. A. Janik, J.Jurkiewicz and M. A. Nowak for stimulating discussions. Special thanks to K. \.{Z}yczkowski for providing us with an algorithm of numerical generation of CUE random matrices.

This work was partially supported by the Polish State Committee for Scientif\mbox{}ic Research (KBN) grant 2P03B08225 (2003--2006). AJ acknowledges the support of the European Network of Random Geometry (ENRAGE) MRTN-CT-2004-005616.



\end{document}